\documentclass[11pt]{article}
\pdfoutput=1
\usepackage{jheppubmod}
\usepackage[punctsep]{collref}
\usepackage{comment}
\collectsep[]{;}
\usepackage{hyperref}

\def\tr{{\rm Tr}}
\def\l{\vec\ell}
\def\x{\vec{x}}
\def\y{\vec{y}}
\def\p{\vec{p}}

\def\ve{\varepsilon}
\def\CG{{\cal G}}
\def\C{{\cal{C}}}
\def\Ru{{\cal{R}}}

\def \q {{\vec{q}}}
\def \P {{\vec{P}}}
\def \k {{\vec{k}}}
\def\Or[#1]{{\text{O}}\left({#1}\right)}
\def\dotl[#1,#2]{\left\langle #1,\, #2 \right\rangle}
\def\dotlb[#1,#2]{\left\langle #1,\, #2 \right\rangle}
\def\dotlm[#1,#2]{\left[ #1,\, #2 \right]}
\def\dotp[#1,#2]{(\vect{#1} \cdot\vect{#2})}
\def\aff[#1,#2]{\hat{#1}(#2)}
\def\n4sym{{\cal N}=4 SYM}
\def\>{\rangle}
\def\<{\langle}
\def\({\left(}
\def\){\right)}
\def\weight[#1,#2,#3]{\{(#1),#2,#3\}}
\def\ads[#1]{$\text{AdS}_{#1}$}

\def\L{{\cal{L}}}

\newcommand{\be}{\begin{equation}}
\newcommand{\ee}{\end{equation}}
\newcommand{\beq}{\begin{eqnarray}}
\newcommand{\eeq}{\end{eqnarray}}
\newcommand{\ba}{\begin{align}}
\newcommand{\ea}{\end{align}}

\newcommand{\tn}{\textnormal}
\renewcommand{\vec}[1]{\boldsymbol{#1}}
\newcommand{\bs}{\begin{split}}
\def\sess\end{split}

\newcommand{\vect}[1]{{\boldsymbol{#1}}}

\usepackage{hyperref}
\usepackage{setspace} 
\title{Onset of many-body chaos in the $O(N)$ model}

\author[1]{Debanjan Chowdhury,}
\affiliation[1]{Department of Physics, Massachusetts Institute of Technology, Cambridge MA
02139, USA.}
\author[1,2,3]{Brian Swingle}
\affiliation[2]{Department of Physics, Harvard University, Cambridge MA
02138, USA.}
\affiliation[3]{Department of Physics, Brandeis University, Waltham MA, 02453, USA.}
\date{\today \\
\vspace{1.6in}}
\abstract{The growth of commutators of initially commuting local operators diagnoses the onset of chaos in quantum many-body systems. We compute such commutators of local field operators with $N$ components in the $(2+1)$-dimensional $O(N)$ nonlinear sigma model to leading order in $1/N$. The system is taken to be in thermal equilibrium at a temperature $T$ above the zero temperature quantum critical point separating the symmetry broken and unbroken phases. The commutator grows exponentially in time with a rate denoted $\lambda_L$. At large $N$ the growth of chaos as measured by $\lambda_L$ is slow because the model is weakly interacting, and we find $\lambda_L \approx 3.2 \,T/N$. The scaling with temperature is dictated by conformal invariance of the underlying quantum critical point. We also show that operators grow ballistically in space with a ``butterfly velocity" given by $v_B/c \approx 1$ where $c$ is the Lorentz-invariant speed of particle excitations in the system. We briefly comment on the behavior of $\lambda_L$ and $v_B$ in the neighboring symmetry broken and unbroken phases.}

\keywords{Scrambling, Quantum chaos, Butterfly effect, Quantum criticality}
\begin{document}
\maketitle

\section{Introduction}

Understanding thermalization in closed quantum systems is one of the great challenges of quantum many-body physics \cite{Deutsch,Srednicki,Tasaki,rigol}. Issues of thermalization have recently received renewed attention, due in part to numerous experiments probing quantum dynamics in approximately closed systems \cite{schmied} \cite{Greiner} \cite{bloch1,bloch2} \cite{marco1,marco2} \cite{monroe1} \cite{Blatt14} as well as new theoretical tools, like AdS/CFT duality, which are able to address dynamics in isolated strongly interaction systems (see Refs. \cite{Chesler15},\cite{SSreview} for a review). In general, there are numerous time scales associated with thermalization, which proceeds through distinct stages. In the simplest scenario, one can identify three distinct stages of thermalization.

At early times there is a process of relaxation, which describes the initial decay of local perturbations. The onset of relaxation can be diagnosed from the time dependence of simple auto-correlation functions of local observables \cite{forster}. The time-scale for relaxation is typically independent of the system size. Relaxation is followed by a process called scrambling at intermediate times \cite{2008JHEP...10..065S}, which describes the spreading of quantum entanglement and information across all of the degrees of freedom of the system. Accessing scrambled information necessarily requires making non-local measurements; the memory of the initial out-of-equilibrium state is effectively lost as far as local probes are concerned \cite{hosur}. For local Hamiltonians the time-scale to reach a scrambled state necessarily depends on the total number of degrees of freedom. Relaxation and scrambling are distinct phenomena, though relaxation processes and time-scales will influence how fast the system scrambles. After the system is effectively scrambled, the quantum state continues to move in Hilbert space and can continue to grow in complexity up to extremely long times \cite{Susskind}.

Scrambling is fundamentally an information theoretic notion \cite{HaydenPreskill}\cite{BrownFawzi}, but the growth of scrambling can also be probed using special correlation functions as we discuss below \cite{ShenkerStanford2014,kitaevtalk}. The relevant correlation functions can be related to information theoretic measures of scrambling, but are easier to access, both theoretically and experimentally. Furthermore, in systems with a semi-classical limit, the growth of scrambling, as measured by these special correlation functions, can be heuristically related to the growth of chaos in the corresponding classical model as measured by classical Lyapunov exponents \cite{kitaevtalk}. However, there are some important subtleties with this connection and it remains incompletely understood \cite{Galitski17}. Because the growth of scrambling describes the effective loss of memory of the initial state and because of its relation to growth of classical chaos, the onset of scrambling can be regarded as a fully quantum avatar of the growth of chaos.

To define the physical correlation functions of interest, consider two local Hermitian operators $W$ and $V$ and a local Hamiltonian $H$. From these objects we can form the time evolution operator $U(t) = e^{-i H t}$ and the Heisenberg operator $W(t) = U(-t) W U(t)$. We take the state of the system to be in thermal equilibrium at temperature $T = 1/\beta$: $\rho \propto e^{-H/T}$. The ``unregulated" squared commutator in state $\rho$ is
\beq
C(t) = \tr\left\{ \rho [W(t),V]^\dagger [W(t),V]\right\} = - \tr\left\{ \rho [W(t),V]^2\right\}
\eeq
where in the second equality we used $[W(t),V]^\dagger = - [W(t),V]$ valid for Hermitian $W$ and $V$. This commutator is called ``unregulated" because some of the operators entering $C(t)$ are inserted at the same spacetime point, a situation which typically leads to additional short-distance divergences in a quantum field theory. The expectation in a quantum chaotic system is that $C(t)$ begins small if $W$ and $V$ are well separated and that it subsequently grows exponentially in time,
\beq \label{sqcommexp}
C(t) \sim \epsilon \,e^{\lambda_L t}
\eeq
for some small parameter $\epsilon$ with an exponent denoted $\lambda_L$. The small parameter $\epsilon$, which in general may also depend on time in some way, is typically related to some measure of the number of relevant degrees of freedom, e.g. the number of components of a field or the distance between $W$ and $V$.

For the calculations in this paper, it is more convenient to consider a ``regulated" squared commutator defined as
\beq
\mathcal{C}(t) = \tr\left\{ \rho^{1/2} [W(t),V]^\dagger \rho^{1/2} [W(t),V] \right\} = -\tr\left\{ \rho^{1/2} [W(t),V] \rho^{1/2} [W(t),V] \right\}.
\eeq
This object has the virtue that all operator insertions occur at distinct spacetime points thus removing short-distance divergences from coincident operators. The regulation of divergences is achieved by moving two of the operators half way around the thermal circle which is equivalent to an analytic continuation off the real time axis.\footnote{To be clear, consider the four point function $\tr\left\{ \rho O_1(t_1) O_2(t_2) O_3(t_3) O_4(t_4) \right\}$. We can move the $j$th operator along the thermal circle by an amount $\tau$ by shifting $t_j \rightarrow t_j - i \tau$ or $O_j(t_j) \rightarrow \rho^{-\tau/\beta} O_j(t_j) \rho^{\tau/\beta}$. Since the thermal circle is periodic with period $\beta=1/T$, $\tau = \beta/2$ corresponds to half way around the thermal circle. Shifting two operators, say $1$ and $2$, half way around amounts to considering $\tr\left\{ \rho^{1/2} O_1(t_1) O_2(t_2) \rho^{1/2} O_3(t_3) O_4(t_4) \right\}$.} We furthermore expect $\mathcal{C}(t)$ to grow exponentially in time at the same rate as $C(t)$. This is because $\mathcal{C}(t)$ is strictly positive, $\mathcal{C} = \tr\left\{ A^\dagger A \right\}$ where $A=\rho^{1/4} [W(t),V] \rho^{1/4}$, so analytically continuing $t$ in $e^{\lambda_L t}$ still yields exponential growth given the guarantee that $\mathcal{C}$ is positive.

\begin{figure}
  \centering
  \includegraphics[width=.4\textwidth]{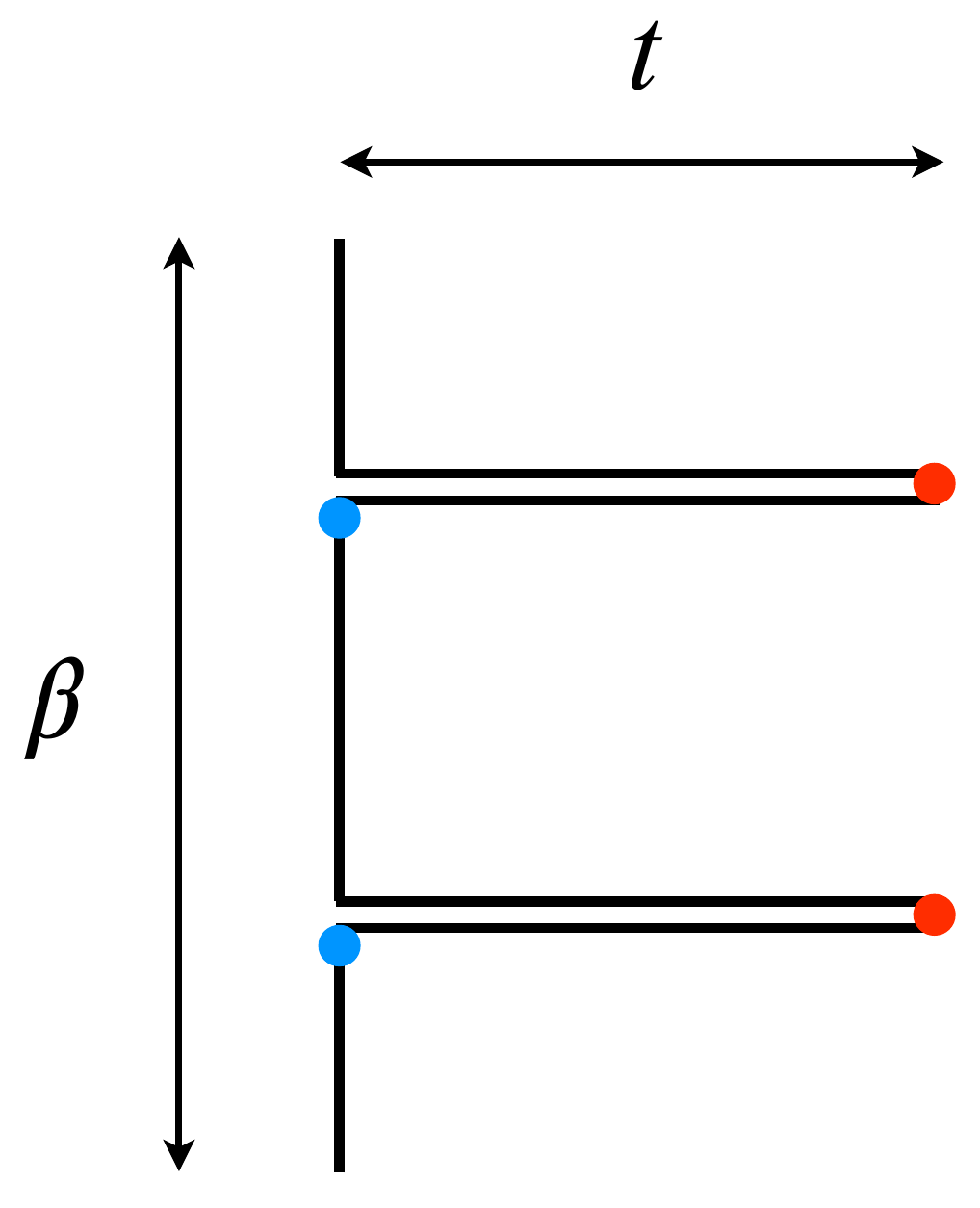}
  \caption{Sketch of the operator ordering in the out-of-time-order correlator $F(t)$, Eq.~\eqref{otodef}. The red dots correspond to $W$ operators and the blue dots correspond to $V$ operators. The two real time folds are separated by an imaginary time of $\beta/2$.}
  \label{otocontour}
\end{figure}

It is also conventional to expand the commutators appearing in $\mathcal{C}(t)$ or $C(t)$ to give two terms:
\begin{eqnarray}
  \mathcal{C}(t) &=& 2\,G(t) - 2\,\text{Re}[F(t)] \\
  G(t) &=& \tr\left\{ \rho^{1/2} W(t) V \rho^{1/2} V W(t) \right\} \\
  \label{otodef} F(t) &=& \tr\left\{ \rho^{1/2} W(t) V \rho^{1/2}  W(t) V \right\}.
\end{eqnarray}
Note that $G$ is positive and time-ordered. On the other hand, $F$ is called an out-of-time-order (OTO) correlator because of the unusual time order of its operator insertions. The placement of the operators in $F$ is shown in Figure~\ref{otocontour}. $F$ is approximately equal to $G$ at early times, when $W(t)$ and $V$ still approximately commute, but differs from $G$ at later times. If $G(t)$ reaches its equilibrium value after some local relaxation time $t_r$, then the subsequent dynamics of $\mathcal{C}(t)$ will be controlled by $F(t)$.

OTO correlators showed up decades ago in the study of semi-classical methods in superconductivity \cite{LO69}. More recently, they have been studied in much greater detail in the context of the AdS/CFT correspondence where they have been shown to diagnose quantum chaos in black hole physics \cite{ShenkerStanford2014} \cite{Shenker2014}\cite{kitaevtalk}. In particular, black holes are supposed to be the fastest scramblers in nature \cite{MSS15}. At the other extreme, disordered quantum systems typically scramble much more slowly than their clean counterparts \cite{BSDC17} \cite{EFDH16} \cite{XC16} \cite{HeLu17} \cite{YC16} \cite{Shen16}. Several experimental proposals have also appeared recently \cite{BSexp} \cite{BSinter} \cite{TGexp} that enable one to measure OTO correlators and scrambling and three preliminary experiments have already been carried out \cite{AMR16} \cite{Du16} \cite{PC16}. There has been a flurry of recent calculations of out-of-time-order correlators in a variety of models \cite{kitaevkitpsyk,2016JHEP...04..001P,2016PhRvD..94j6002M,BSLR,DSweak,BanerjeeAltman,PatelSachdev,2016arXiv161200614D,Aleiner16,Knap16,2017arXiv170203929L,2016arXiv161002669L,FKmodel}.
There have also been some recent works exploring other formal aspects of these correlators, including fluctuation-dissipation-like theorems \cite{nyhOTOC}\cite{2016arXiv160905848C}\cite{2016arXiv161208781T}\cite{2017arXiv170102820H}.

In this paper we study scrambling in a class of quantum critical field theories, namely vector models associated with an $N$ component real scalar $\varphi_a$ field in (2+1)-dimensions with $O(N)$ symmetry. These theories describe a zero-temperature quantum phase transition from a disordered phase where $\langle \varphi_a \rangle = 0$ to an ordered symmetry breaking phase where $\langle \varphi_a \rangle \neq 0$. A large class of problems at long wavelengths are described by this model, e.g. the theory with $N=2$ describes the transition between a superfluid and a bosonic Mott insulator \cite{Fisher89} and the theory with $N=3$ describes the transition from a paramagnet to a Heisenberg antiferromagnet \cite{CHN89} \cite{SY92} \cite{CSY94}. While the system can be described by well-defined quasiparticles on either side of the transition, the critical point itself is distinguished by the absence of any quasiparticle-like excitations. Moreover, the quantum many-body system at a non-zero temperature `above' the quantum critical point (QCP) \cite{QPT} is described by only one relevant scale, namely the temperature itself. It is then natural to ask how fast this broad class of theories scrambles at a given temperature.

Before stating our results, we will explicitly describe the model. We study the theory of an $O(N)$ symmetric real vector field, $\varphi_a$, $a=1,..,N$, governed by the real time Lagrangian
\beq
\L =  \bigg[\frac{1}{2}(\partial\varphi_a)^2 - \frac{v}{2N}\bigg(\varphi_a^2 - \frac{N}{g}\bigg)^2 \bigg].
\label{on}
\eeq
The real time generating functional is
\beq
\tilde{Z}=\int {\cal{D}}\varphi_a~ \tn{exp}\bigg(i\int_x \L\bigg),
\eeq
where $\int_x\equiv\int dt~d^2\x$ is the integral over $(2+1)-$ dimensional space-time{\footnote{We reserve $\tau ~(i\omega_n) $ for  Euclidean-imaginary time (Matsubara frequency) and $t~ (\omega)$ for real time (frequency).}}. The parameter $g$ is used to tune across the QCP (which occurs at $g=g_c$) and $v>0$ is a self-interaction coupling constant. We have set the ``speed of light" $c=1$.

We study the index-averaged commutator of the fields, $\varphi_a(\x,t)$,
\beq
\mathcal{C}(t,\x) = - \frac{1}{N^2} \sum_{a,b} \tr\left\{\rho^{1/2} ~[\varphi_a(\x,t),\varphi_b] \rho^{1/2} ~[\varphi_a(\x,t), \varphi_b] \right\},
\label{ct0}
\eeq
where $\varphi_b = \varphi_b({\vec 0},0)$ and $\rho = e^{-\beta H}/Z$ is the thermal density-matrix. This corresponds to the thermally regulated commutator discussed above with $W = \varphi_a(\x)$ and $V = \varphi_b({\vec 0})$. The normalization in Eq.~\eqref{ct0} amounts to averaging the square commutator over $a,b=1,...,N$.

Our main results are summarized below. All results are to leading non-trivial order in a $1/N$ expansion at the quantum critical point at temperature $T$. We first compute the chaos exponent $\lambda_L$ in the spatially averaged case, i.e. the growth rate of $\int d^2 \x ~\mathcal{C}(t,\x)$, and find
\beq
\lambda_L \approx 3.2 \frac{T}{N}.
\eeq
As discussed below, our calculation of the prefactor is approximate; we roughly estimate the accuracy to be about $10$\%.

 We next analyze the commutator, $\mathcal{C}(t,\vec{k})=\int d^2 \x ~e^{-i\k\cdot\x}\mathcal{C}(t,\x)$, at a fixed external momentum, $\vec{k}$. This grows exponentially in time as well with a $\vec{k}$ dependent growth exponent given by
\beq
\lambda(\vec{k}) = \lambda_L - D_L | \vec{k}|^2 + ...
\eeq
The spacetime structure of the commutator is determined from the Fourier transform
\beq
\C(t,\x) \sim \int_\k \chi_\k~ e^{i\k\cdot\x + \lambda_L(\k) t}
\label{poles}
\eeq 
which we evaluate in a saddle point approximation (assuming no singular structure in $\chi_\k$). In holographic and related studies \cite{StanfordStringy,GuQiStanford}, there is an additional pole-like singularity in $\chi_\k$ which, when closer to the real-axis, determines the spatial decay. As we discuss later, we see no evidence for such a pole in our numerical computations and speculate that it may be related to the presence of gravitational modes.

Within the above approximation, the space-time dependence of $\mathcal{C}(t,\x)$ then takes the form
\beq \label{diffusion}
\mathcal{C}(t,\x) \sim \exp\left(\lambda_L t - \frac{|\x|^2}{4 D_L t}\right)
\eeq
where $\lambda_L$ is the uniform growth exponent and $D_L$ has units of a diffusion constant. Comparing Eq.~\eqref{diffusion} to Eq.~\eqref{sqcommexp} shows that the role of $\epsilon$ is played by
\beq
\epsilon \sim \exp\left(- \frac{|\x|^2}{4 D_L t} \right),
\eeq
which for large $|\x|$ indeed represents a suppression factor in the square commutator. 

From Eq.~\eqref{diffusion} it follows that $\mathcal{C}(t,\vec{x})$ obeys a local equation of motion of the form
\beq
\partial_t \mathcal{C}(t,\vec{x}) = \lambda_L \mathcal{C}(t,\vec{x}) + D_L \nabla^2 \mathcal{C}(t,\vec{x}) + ...
\eeq
where $...$ denotes higher order terms in a gradient expansion and non-linear terms. In practice, we directly compute the momentum dependent chaos exponent and only obtain the spacetime structure indirectly.

The form in Eq.~\eqref{diffusion} is consistent with a ballistic growth of operators with a ``butterfly velocity" \cite{RSS15}
\beq
v_B = \sqrt{4 D_L \lambda_L}.
\eeq
We find that $v_B$ is independent of $T$ and $N$ at large $N$; expressed as a ratio with the speed of light, our approximate numerical calculation gives
\beq
\frac{v_B}{c} \approx 1.
\eeq
As discussed below, our calculation of the ratio is approximate; we roughly estimate the accuracy to be about $10$\%. Of course, causality constrains $v_B/c$ to be less than one. The butterfly velocity has been argued \cite{BSLR} to play the role of a low energy state dependent Lieb-Robinson velocity \cite{LiebRobinson}.

Interestingly, the derivative of the diffusive-plus-growth form in Eq.~\eqref{diffusion} is
\beq
\frac{1}{\mathcal{C}}\frac{d\mathcal{C}}{dt} = \lambda_L + \frac{|\x|^2}{4 D_L t^2},
\eeq
which for sufficiently large $|\x|/t$ is greater than the chaos bound of $2\pi T$ \cite{MSS15}. Of course, the chaos bound does not directly address the dynamics of $\mathcal{C}$, but the bound does suggest that the diffusive form must break down for sufficiently large $|\x|/t$. For example, higher order terms in the $\k$ expansion of $\lambda(\k)$ will become important as the saddle point momentum increases away from the real axis. Additional singularities, e.g. poles and branch cuts, and the exact vanishing of the commutator outside the lightcone may also become relevant before the chaos bound is violated. 

There is another feature of note concerning the growth of chaos in space. Expanding Eq.~\eqref{diffusion} near the wavefront $|\x| = v_B t$ gives
\beq
\mathcal{C}(t,\x)\big|_{|t - |\x|/v_B| \ll t} \sim \exp\left(2 \lambda_L [t - |\x|/v_B] \right).
\eeq
Note the peculiar factor of $2$ in front of $\lambda_L$ which arises from the expansion. The full behavior of the squared commutator near the front is not fully understood at this point and is beyond the scope of this work. Moreover, the nature of the crossover of the squared commutator outside of the causal region will be the topic of future work.

\begin{figure}
  \centering
  \includegraphics[width=0.7\textwidth]{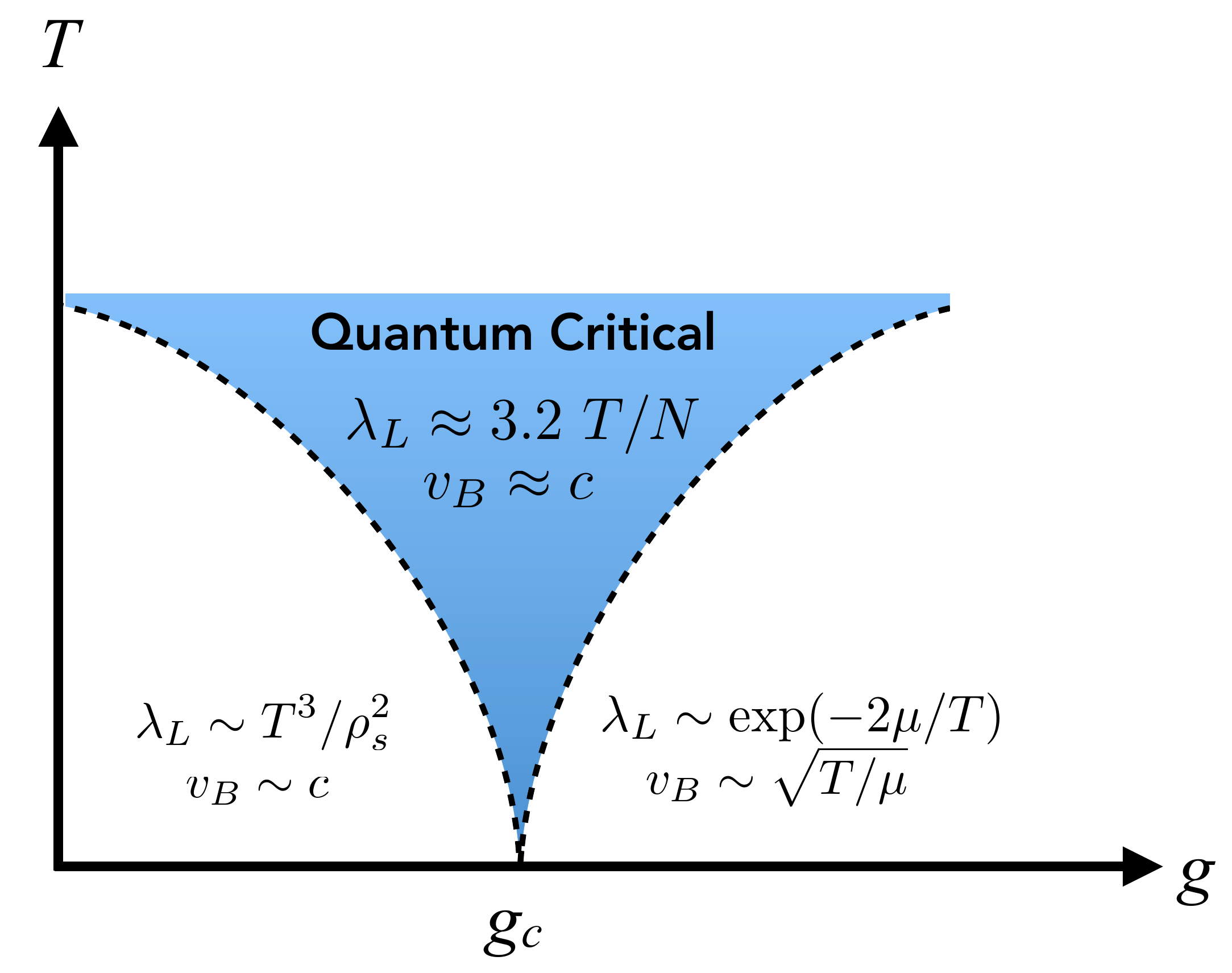}
  \caption{A summary of the results for the growth exponent, $\lambda_L$, and butterfly velocity, $v_B$, in the different regions of the $g-T$ phase diagram (see Eq. \eqref{on}). This paper is primarily concerned with the `quantum critical' regime (blue shaded region) where temperature is the only relevant energy scale. The results for the symmetry-broken (unbroken) regions, corresponding to $g<g_c$ ($g>g_c$), are discussed briefly in Section \ref{proximate}.}
  \label{phase}
\end{figure}

We also estimate $\lambda_L$ and $v_B$ in the proximate phases. In the symmetry unbroken phase at low temperature we find $\lambda_L \sim e^{-2\mu/T}$ ($\mu$ is the mass of the $\varphi$ particles) and $v_B \sim \sqrt{T/\mu}$. In the symmetry broken phase at low temperature we find $\lambda_L \sim T^3/\rho_s^2$ and $v_B/c \sim 1$. A word of caution is appropriate here: in two spatial dimensions at finite $N$ and non-zero temperature the symmetry broken phase is disordered and the symmetry is restored. Our estimates are based on the existence of a Goldstone mode which will not persist at the longest distances at finite temperature. We discuss this subtlety more carefully in Section \ref{proximate}. Our results for the quantum critical region and our estimates for the proximate phases are summarized in Figure \ref{phase}; in general, there is a complex crossover function which connects the three regimes discussed above.

The rest of this paper is organized as follows: in Section \ref{prelim}, we set the stage for carrying out a $1/N$ expansion for the $O(N)$ theory described by Eq.~\eqref{on}. Section \ref{ladder} develops the perturbative expansion used to compute the regulated commutator $\mathcal{C}(t,\vec{x})$. Sections \ref{LE} and \ref{VB} discuss the numerical solution of the ladder series derived in Section \ref{ladder} to obtain the growth exponent and the butterfly velocity. We briefly discuss scrambling in the symmetry broken and unbroken phases in Section \ref{proximate}, leaving a detailed computation for future work. We end with a discussion where we highlight some interesting future directions. The appendices contain a variety of technical details related to our calculations.\\ \\
{\bf Notation guide}: In the remainder of the paper, $N$ denotes the number of components of the field $\varphi_a$ and $\lambda$ denotes the Lagrange multiplier in the $O(N)$ action. The imaginary time $\varphi$ correlator is $\CG$, the retarded $\varphi$ correlator is $\CG_R$, and the Wightman $\varphi$ correlator is $\CG_W$. Similarly, the imaginary time $\lambda$ correlator is $\CG_\lambda$, the retarded $\lambda$ correlator is $\CG_{R,\lambda}$, and the Wightman $\lambda$ correlator is $\CG_{W,\lambda}$. The thermal mass of the $\varphi$ field is denoted $\mu$ and $\Lambda$ is a momentum cutoff where needed. $\lambda_L$ is the chaos exponent and $v_B$ is the butterfly velocity. We distinguish position and time and momentum and frequency with spatial vectors denoted by bold face, e.g. $\vec{x}$ and $\vec{k}$. Integrals over momentum are denoted $\int_{\vec{k}}$ and include the $(2\pi)^{-2}$ phase space factor in $d=2$ spatial dimensions. The thermal state is $\rho \propto e^{-\beta H}$ where $H$ is the Hamiltonian and $\beta = 1/T$ is the inverse temperature. We set the speed of light $c=1$, unless explicitly stated.

\section{Preliminaries}
\label{prelim}

In this and the next section, we set up the formalism required to compute $\mathcal{C}(t)$ in a $1/N$ expansion in the theory Eq.~\eqref{on}. We are particularly interested in the limit of strong coupling $v\rightarrow \infty$ at the quantum critical point where Eq.~\eqref{on} describes a $2+1$-dimensional conformal field theory \cite{QPT}.

Decoupling the quartic term in Eq.~\eqref{on} using a Hubbard-Stratonovich field, $\lambda(\x,t)$, leads to a new Lagrangian
\beq
\L\rightarrow \overline{\L} = \frac{1}{2}\bigg[(\partial\varphi_a)^2 + \frac{\lambda}{\sqrt{N}} \bigg(\varphi_a^2 - \frac{N}{g}\bigg) + \frac{\lambda^2}{4v} \bigg],
\label{HSl}
\eeq
such that the generating function becomes
\beq
\tilde{Z}=\int {\cal{D}}\varphi_a~{\cal{D}}\lambda~\tn{exp}\left(i\int_x \overline\L\right).
\eeq

Although we are ultimately interested in real time evolution, it is also quite useful to consider the imaginary time formulation of the above theory. As usual we set $i t \rightarrow \tau$ and construct the Euclidean Lagrangian
\beq
\L_E = \frac{1}{2}\bigg[(\partial\varphi_a)^2 - \frac{\lambda}{\sqrt{N}} \bigg(\varphi_a^2 - \frac{N}{g}\bigg) - \frac{\lambda^2}{4v} \bigg]
\eeq
and the partition function
\beq
Z = \int {\cal{D}}\varphi_a~{\cal{D}}\lambda~\tn{exp}\left(- \int d\tau d^2 \x \, \L_E \right).
\eeq
The derivative term now uses the Euclidean $+++$ metric instead of the $+--$ metric in real time.\footnote{Note that for comparison with some of the literature, the field $\lambda$ is sometimes traded for the field $\tilde{\lambda} = i \lambda$. We will work with $\lambda$ consistently throughout because it is naturally regarded as Hermitian.}

To leading order in $N$, the effect of $\lambda$ is simply to generate an effective mass for the particles described by the $\varphi$ field: $-\langle \lambda \rangle/\sqrt{N} = \mu^2$. The imaginary time $\varphi_a$ propagator is
\beq
\CG(\tau,\x)\delta_{a,b} = \langle \varphi_a(\x,\tau) \varphi_b({\vec 0},0) \rangle.
\eeq
In momentum space $\CG$ has the simple form,
\beq
\CG(i \omega_n,\k) &=& \frac{1}{\omega_n^2 + \epsilon^2_\k},~\tn{where}\\
\epsilon_\k^2 &=& \k^2 + \mu^2.
\label{g1}
\eeq
It is important to note that at the critical point $g=g_c$, the mass term $\mu$ for $\varphi_a$ is proportional to the temperature since there are no other scales. The mass, $\mu = \Theta T$, where $\Theta = 2\log \frac{1+\sqrt{5}}{2}$; the standard computation is reviewed in Appendix \ref{mass}.

The retarded propagator is defined in the usual way as
\beq
\CG_R(t,\vec{x}) \delta_{a,b} = - i \langle [\varphi_a(\vec{x},t),\varphi_b(\vec{0},0)]\rangle \theta(t)
\eeq
where $\theta(t)$ is the Heaviside step function. $\CG_R$ is related to $\CG$ by analytic continuation,
\beq
\CG_R(\omega,\k) = -\CG(i\omega_n \rightarrow \omega + i 0^+,\k) = \frac{1}{(\omega + i 0^+)^2 - \epsilon^2_\k},
\eeq
and the spectral function $A(\omega,\k) = - 2 \,\text{Im}[\CG_R(\omega,\k)]$ is
\beq
A(\omega,\k) = \frac{\pi}{\epsilon_\k}[ \delta(\omega - \epsilon_\k) - \delta(\omega + \epsilon_\k)].
\eeq
To fix notation, recall that the inverse formula is
\beq
\CG_R(\omega,\k) = \int \frac{d\nu}{2\pi} \frac{A(\nu,\k)}{\omega + i0^+ - \nu}.
\eeq

At infinite $N$, this is the whole story. The $\varphi$ particles are massive but non-interacting and chaos does not develop. In particular, the squared commutator is
\beq
\mathcal{C}_0(t,\x) = \left[\CG_R(t,\x)\right]^2.
\eeq
This implies that the chaos exponent is of order $1/N$ at large $N$. Hence we must study $1/N$ corrections in the theory Eq.~\eqref{HSl}.

In order to evaluate the squared commutator to higher orders in $1/N$, we need additional real-time propagators. Thus we also define the symmetrized Wightman function
\beq
\CG_W(t,\x) \delta_{a,b} = ~\tn{Tr}\left\{\rho^{1/2}~ \varphi_a(\x,t) ~\rho^{1/2}~ \varphi_b(0) \right\}.
\eeq
As reviewed in Appendix \ref{wight}, this function may be written in terms of the corresponding $\varphi$ spectral function as
\beq
\CG_W(\omega,\k) =  \frac{A(\omega,\k)}{2 \sinh \frac{\beta \omega}{2}}.
\label{wightspec}
\eeq

\begin{figure*}
  \centering
  \includegraphics[width=.7\textwidth]{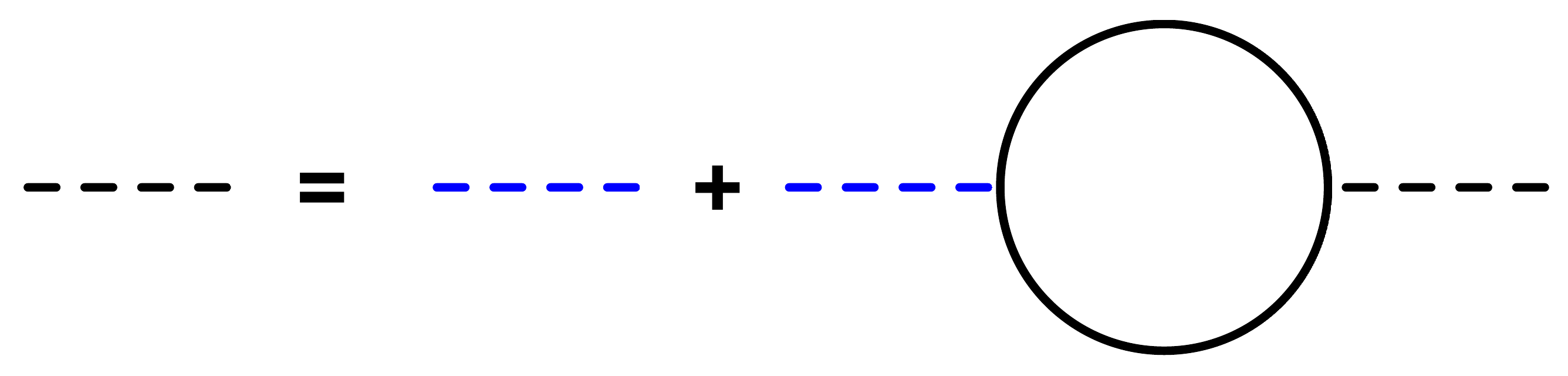}
  \caption{Resummed bubble diagrams, $\Pi(i\omega_n,\k)$, renormalize the bare $\lambda$ propagator $\CG_\lambda^{(0)}(i\omega_n,\k)$  (blue dashed line) to yield $\CG_\lambda(i\omega_n,\k)$ (black dashed line).}
  \label{bubble}
\end{figure*}

For our subsequent computations involving $1/N$ corrections, we also need the propagator for the $\lambda$ field.  The bare Euclidean propagator is $\CG^{(0)}_\lambda(i\omega_n,\k)=-4v$. At infinite $v$, we may re-sum and dress the $\lambda$ propagator as shown in Figure~\ref{bubble}. Then,
\beq
\CG_\lambda(i\omega_n,\k) = \frac{\CG_\lambda^{(0)}}{1-\Pi~\CG_\lambda^{(0)}} \underbrace{\longrightarrow}_{v \rightarrow \infty} - \frac{1}{\Pi(i\omega_n,\k)},
\eeq
where $\Pi(i\omega_n,\k)$ is the one-loop $\varphi_a$ bubble,
\beq
\Pi(i \omega_n,\q) = \frac{T}{2}~\sum_{i\nu_n} \int^\Lambda_\k \frac{1}{(\nu_n + \omega_n)^2 + \epsilon_{\k+\q}^2} ~\frac{1}{\nu_n^2 + \epsilon^2_\k}.
\label{piexp}
\eeq

We evaluate the above diagram in Appendix \ref{oneloop}. We define the retarded bubble $\Pi_R(\omega,\q) = \Pi(i\omega_n \rightarrow \omega + i0^+,\q)$. The real and imaginary parts of the retarded bubble $\Pi_R(\omega,q)$ are evaluated numerically as discussed in Appendix \ref{sec:numerics} and serve as inputs to the later calculation of $\lambda_L$ and $v_B$.

From the basic Euclidean $\lambda$ propagator we extract the real time correlators of the $\lambda$ field. First, the retarded $\lambda$ correlator is
\beq \label{lambdaeuctoret}
\CG_{R,\lambda}(\omega,\k) = - \left( \frac{-1}{\Pi(i\omega_n \rightarrow \omega + i0^+,\k)}\right) = \frac{1}{\Pi_R(\omega,\k)}.
\eeq
The $\lambda$ spectral function is then
\beq
A_\lambda(\omega,\k) = - 2 ~\text{Im}\left[\frac{1}{\Pi_R(\omega,\k)} \right] = \frac{2~ \text{Im}[\Pi_R]}{\text{Im}[\Pi_R]^2 + \text{Re}[\Pi_R]^2}.
\eeq
The imaginary part of the retarded bubble is positive for positive frequency, so the $\lambda$ spectral function is also positive for positive frequency. From this spectral function we immediately obtain the symmetrized Wightman function for $\lambda$,
\beq
\CG_{W,\lambda}(\omega,\k) = \frac{A_\lambda(\omega,\k)}{2\sinh \frac{\beta \omega}{2}}.
\eeq
The $\lambda$ Wightman function is also positive.

Finally, in order to carry out the computation consistently to leading order in $1/N$, we need to include the leading $1/N$ correction to the self-energy for the $\varphi_a$ propagator (Figure~\ref{self}). This is given by,
\beq
\Sigma(i \omega_n, \q) &=& \tilde\Sigma(i \omega_n, \q) + \frac{1}{\Pi(0,\vec{0})} T \sum_{i\nu_n} \int_\k^\Lambda ~[\CG(i\nu_n,\k)]^2~  \tilde\Sigma(i \nu_n, \k),\\
\tilde\Sigma(i \omega_n, \q) &=& \frac{T}{N}  \sum_{i\nu_n} \int^\Lambda_\k ~\bigg[ \CG(i \omega_n + i \nu_n, \q+\k) - \CG(i\nu_n,\k)\bigg] ~\CG_\lambda(i \nu_n,\k).
\label{self}
\eeq
The two contributions in the first line above correspond to the two diagrams in Figure~\ref{self}.
The self-energy modifies the $\varphi$ propagator,
\beq
\CG(i\omega_n,\q)= \frac{1}{\omega_n^2 + \epsilon_\q^2 - \Sigma(i\omega_n,\q)}.
\eeq

As with the bubble, we define the retarded self-energy by $\Sigma_R(\omega,\q) = \Sigma(i \omega_n \rightarrow \omega+i0^+,\q)$. Converting the self-energy modified Euclidean propagator to the corresponding retarded correlator yields
\beq
\CG_R(\omega,\q) = \frac{1}{(\omega+i0^+)^2 - \epsilon_\q^2 + \Sigma_R(\omega,\q)}.
\eeq
For our purposes, the important physical effect of $\Sigma_R$ is to introduce a finite lifetime to the $\varphi$ particles. The poles of $\CG_R$ are shifted by $\Sigma_R$, so working to first order in $\Sigma_R$ gives poles at
\beq
\omega = \omega_* = \pm \epsilon_\q \left( 1 - \frac{\Sigma_R(\pm \epsilon_\q,\q)}{2 \epsilon^2_\q} \right).
\eeq
Since the real part of $\Sigma_R$ is symmetric and the imaginary part is anti-symmetric, we find the poles
\beq
\omega_* = \pm \left(\epsilon_\q - \frac{\text{Re}[\Sigma(\epsilon_\q,\q)]}{2 \epsilon_\q} \right) - i \frac{\text{Im}[\Sigma_R(\epsilon_\q,\q)]}{2 \epsilon_\q}.
\eeq
This suggests defining the inverse lifetime
\beq
\Gamma_\q = \frac{\text{Im}[\Sigma_R(\epsilon_\q,\q)]}{2 \epsilon_\q},
\eeq
so that $\Gamma_\q > 0$ corresponds to decay and $\CG_R$ remains analytic in the $\omega$ upper half-plane.

We have evaluated the imaginary part of the self-energy in Appendix \ref{SE}; the final expression is given by,
\beq
\tn{Im}[\Sigma_R(\omega,\q)] = \frac{1}{N}\int^\Lambda_\k \frac{\sinh(\beta\omega/2)}{4\epsilon_\k\sinh(\beta\epsilon_\k/2)}\bigg[\CG_{W,\lambda}(\epsilon_\k-\omega,\k-\q) + \CG_{W,\lambda}(-\epsilon_\k-\omega,\k-\q)\bigg].\nonumber\\
\label{seg}
\eeq
As a check on the computation, notice that positivity of the $\lambda$ Wightman function implies positivity of the decay rate $\Gamma_{\vec{q}}$.

\begin{figure}
  \centering
  \includegraphics[width=.7\textwidth]{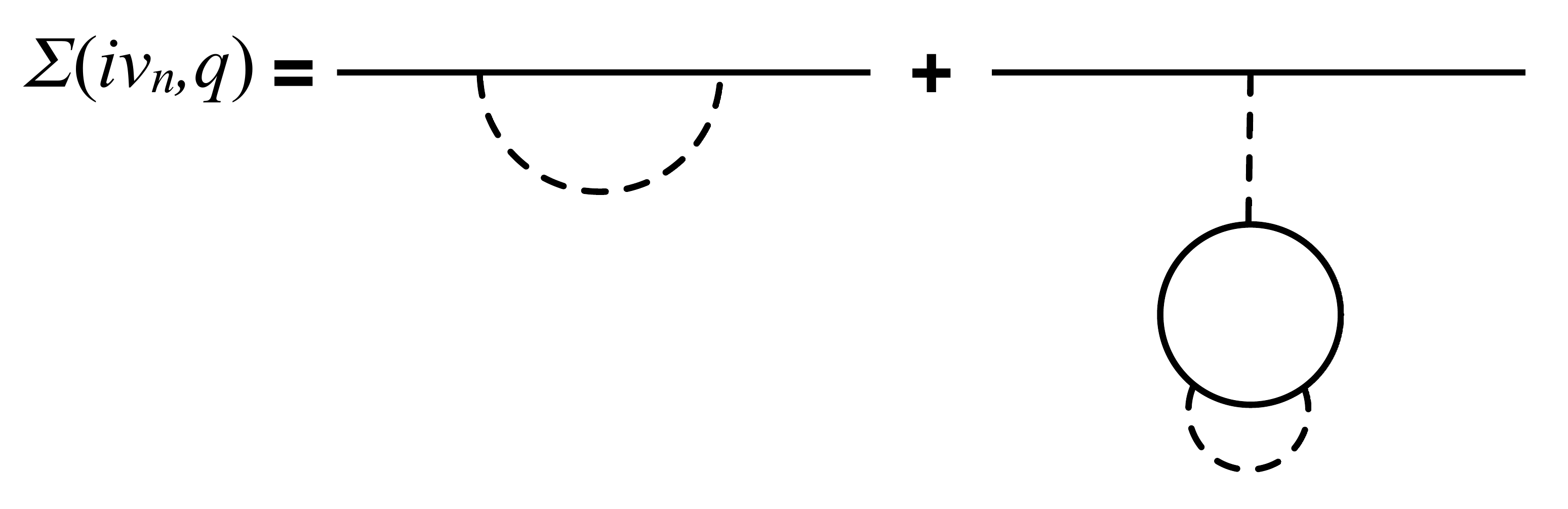}
  \caption{The self-energy corrections to the $\varphi_a$ propagator. The second diagram is independent of the external momentum/frequency.}
  \label{self}
\end{figure}

The inverse lifetime is
\beq
\Gamma_\q &=&  \frac{1}{2 N}\int^\Lambda_\k \frac{\sinh(\beta\epsilon_\q/2)}{4\epsilon_\k\epsilon_\q\sinh(\beta\epsilon_\k/2)}\bigg[\CG_{W,\lambda}(\epsilon_\k-\epsilon_\q,\k-\q) + \CG_{W,\lambda}(-\epsilon_\k-\epsilon_\q,\k-\q)\bigg],\nonumber\\
\Gamma_\q &=& \frac{1}{2 N}\int^\Lambda_\k \frac{\sinh(\beta\epsilon_\q/2)}{\sinh(\beta\epsilon_\k/2)} \Ru_1(\k,\q),
\label{gammaq}
\eeq
where the function $\Ru_1(\k,\q)$ is given by,
\beq
\Ru_1(\k,\q) &=&  R_{1,+}(\k,\q) + R_{1,-}(\k,\q),\\
R_{1,\pm}(\k,\q) &=&  \frac{\CG_{W,\lambda}(\pm\epsilon_{\k} - \epsilon_{\q},\k-\q)}{4\epsilon_{\k}\epsilon_{\q}}.
\eeq
The significance of $\Ru_1$ will become clear shortly.

We can define an inverse ``(on-shell) phase-coherence time",
\beq
\frac{1}{\tau_\varphi} = \Gamma_{\vec{q}=0} = \frac{\sinh(\beta\mu/2)}{2 N}\int^\Lambda_\k \frac{1}{\sinh(\beta\epsilon_\k/2)} \Ru_1(\k,\vec{0}).
\eeq
$\tau_\varphi$ is associated with the loss of phase coherence and can be deduced from a two-point function. It is the analog of the relaxation time discussed in the introduction. We find
\beq
\frac{1}{\tau_\varphi} = 1.152 \frac{T}{N}.
\eeq
This is a high precision value good to three digits{\footnote{We thank W. Witczak-Krempa and S. Sachdev for helping benchmark the numerical computation of $\tau_\phi$ \cite{WWKSS12}.}} which serves as a check on our numerical calculation of $\Pi$ and as a benchmark for the more complex calculations of $\lambda_L$ and $v_B$ since these calculations also output approximate values for $1/\tau_\varphi$.

\section{Diagrammatic expansion for the squared commutator}
\label{ladder}

We now assemble the pieces discussed in the previous sections to compute $\mathcal{C}(t)$ in theory Eq.~\eqref{HSl} to leading non-trivial order in $1/N$. Recall that as shown in Figure~\ref{otocontour} the squared commutator is naturally associated with a complex time contour which includes the imaginary time thermal circle and two ``real time folds". The perturbation expansion is naturally organized in terms of retarded correlation functions within the time folds and Wightman correlation functions between the two time folds.

To generate the perturbation series, we expand $\mathcal{C}$ in powers of the interaction vertex on both time folds. As far as possible, we keep only diagrams which affect the result at order $1/N$. However, we re-sum a number of contributions, so we are not performing a strict $1/N$ expansion since diagrams with arbitrarily high powers of $1/N$ are included. Some additional classes of diagrams which are not $1/N$ suppressed are nevertheless ignored, as discussed below, because they do not affect the leading $1/N$ growth rate.

\begin{figure}
  \centering
  \includegraphics[width=.8\textwidth]{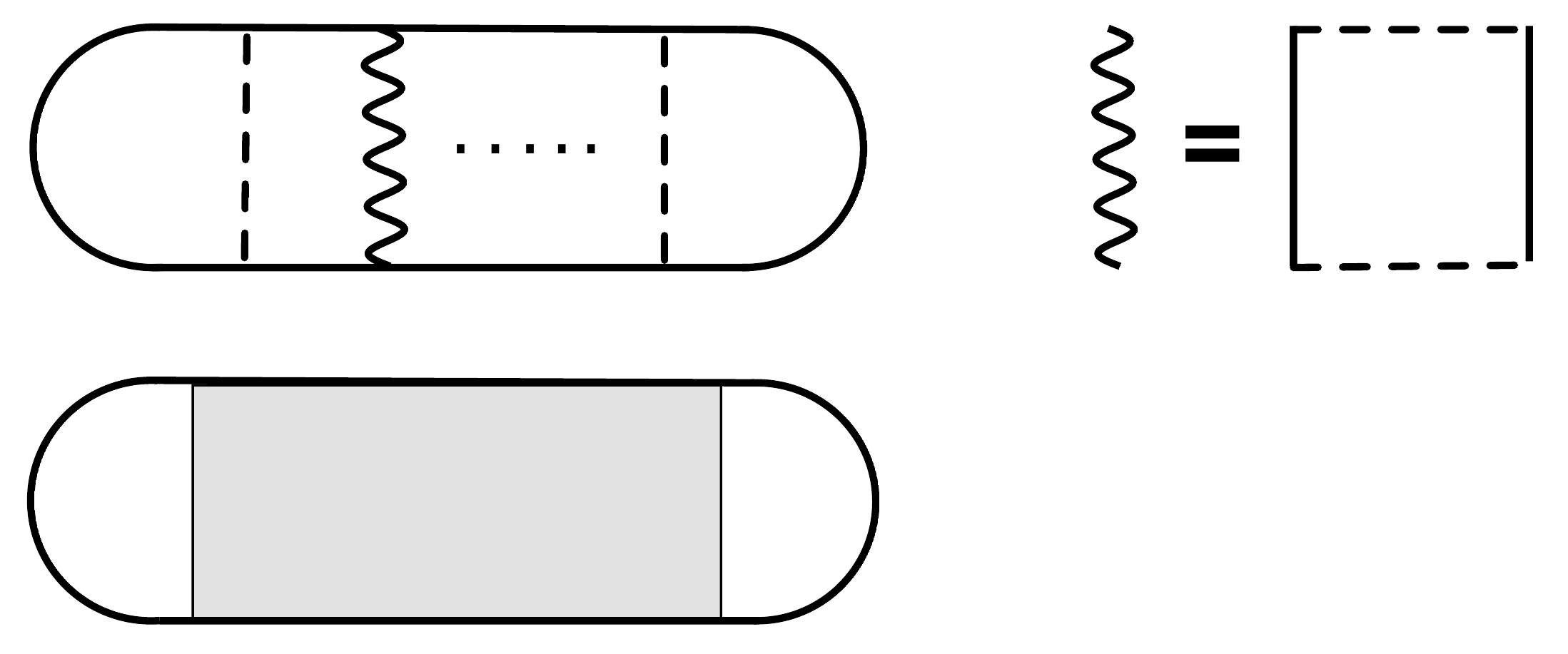}
  \caption{The top diagram represents a general uncrossed ladder diagram in which two kinds of rungs are allowed. The first rung (dashed black line---type-I rung) corresponds to the insertion of $\CG_{W,\lambda}$ between the two sides of the ladder. The second rung (wavy black line---type-II rung) corresponds to the insertion of the box diagram shown at right into the ladder. The contribution of the box is denoted $\CG_{\text{eff}}$ and is defined in Eq.~\eqref{Geff}. The bottom diagram represents the ladder sum over the two kinds of rungs, for which we write down a Bethe-Saltpeter equation in Eq.~\eqref{resummed} (see Figure \ref{ladsum}).}
  \label{laddersum}
\end{figure}

The rules of our calculation are discussed in detail in Appendix \ref{sec:diagramrules}. They are summarized as follows:
\begin{enumerate}
\item Vertex insertions are restricted to lie along the real time folds. Each real time vertex insertion is associated with a factor of $\frac{i}{2\sqrt{N}}$. Vertex insertions along the thermal circle dress the thermal state but do not directly lead to real time growth; the relevant effects are included in the various finite temperature correlation functions obtained via analytic continuation from imaginary time. Mixed insertions involving contractions between operators on the thermal circle and operators on a time fold also do not grow in time. In fact, they decay in time as the operator on the time fold moves away from the thermal circle.
\item Horizontal lines correspond to retarded propagators, $i \CG_R$ or $i \CG_{R,\lambda}$, and vertical lines correspond to Wightman propagators, $\CG_W$ or $\CG_{W,\lambda}$. The direction of the lines is not meaningful in Euclidean diagrams. The Euclidean $\lambda$ propagator is given by $-1/\Pi$ to leading order.
\item The fastest growing diagrams correspond to a set of ladder diagrams with two types of rungs as shown in Figure~\ref{laddersum}. The type-I rung (dashed line) corresponds to $\CG_{W,\lambda}$ while the type-II rung (wavy line) corresponds to the box insertion which is schematically $\sim\CG_{R,\lambda}^2 \CG_W^2$. When considering the averaged squared commutator, the sum over $a$ and $b$ effectively caps off the ladder as shown in Figure~\ref{laddersum}.
\item Accounting for the appropriate closed loops of $\varphi$ which appear when considering the type-II rung, both rungs are associated with a factor of $1/N$. This leads to a factor of $1/N^\ell$ for a diagram with $\ell$ rungs. Hence $\lambda_L \sim 1/N$.
\item In addition to the explicit rungs, the $\varphi$ sides of the ladder are dressed with self-energy corrections. The self-energy is computed to order $1/N$.
\item Vertex corrections and other higher order diagrams are ignored. Crossed ladder diagrams are of the same order in $1/N$ as the ladder diagrams we keep, but we argue in Appendix \ref{cross} that these diagrams do not contribute to the leading order growth rate. Roughly speaking, this is because they are far from being on-shell.
\end{enumerate}

A few comments are in order. The focus on ladder diagrams is the standard perturbative approach to the calculation of chaos exponents \cite{kitaevkitpsyk,2016JHEP...04..001P,DSweak,2016PhRvD..94j6002M}. Our calculation extends a recent perturbative calculation in a $\varphi^4$ matrix-model in which the fields $\varphi$ are $N \times N$ matrices \cite{DSweak}. There are some important differences between the vector and matrix models. The growth exponent is $1/N$ suppressed in the vector model but not the matrix model due to differences in the $N$ counting. The crossed ladder diagrams are $1/N$ suppressed in the matrix model because they are non-planar. The type-II rungs are also suppressed due to the weak-coupling treatment of the matrix model. There are additional re-summations in the vector model, e.g. to compute the renormalized $\lambda$ propagator, and we are able to access a conformal fixed point in the vector model.

The explicit $N$ counting is straightforward. A ladder with $\ell_1$ rungs of type-I and $\ell_2$ rungs of type-II, denoted $D_{\ell_1,\ell_2}$, is associated with a factor of
\beq
D_{\ell_1,\ell_2} \sim \underbrace{\frac{1}{N^2}}_{\text{Def}} \times \underbrace{N^{\ell_2+1}}_{\text{Loops}} \times \underbrace{\(\frac{1}{\sqrt{N}}\)^{2\ell_1 + 4\ell_2}}_{\text{Vertices}} = \frac{1}{N^{\ell_1+\ell_2+1}}.
\eeq
Hence the $N$ suppression of a ladder depends only on the total number of rungs of either type, $\ell = \ell_1 +\ell_2$. Summing over all ladder diagrams then gives a square commutator of order
\beq
\mathcal{C} \sim \sum_\ell D_\ell \sim \frac{1}{N}.
\eeq
We also see that the growth rate will have an explicit factor of $1/N$. The scrambling time, $t_*$, defined as $\mathcal{C}(t_*)\sim 1$ will thus be of order
\beq
t_* \sim \beta N \ln N.
\eeq
This is consistent with earlier arguments \cite{Perlmutter2016}.

\begin{figure}
  \centering
  \includegraphics[width=.8\textwidth]{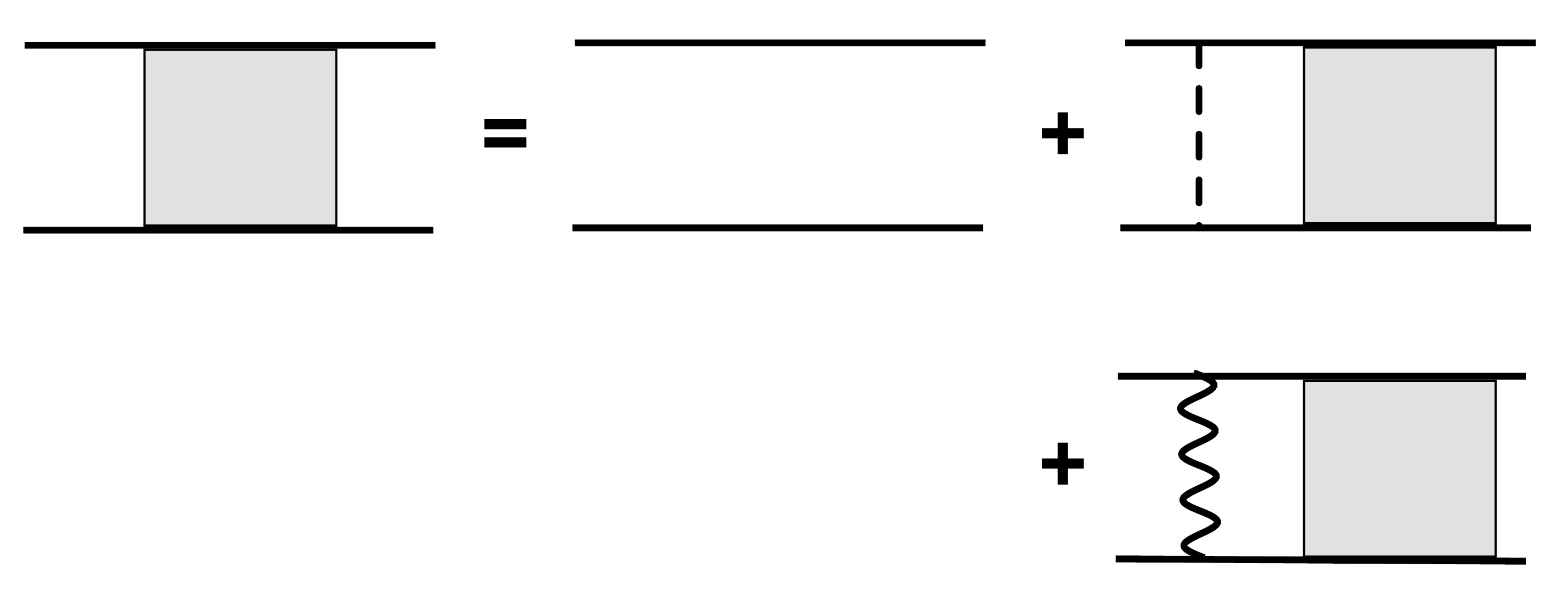}
  \caption{Bethe-Saltpeter equation for the growing piece of $f$ (or equivalently, $\mathcal{C}$). The first diagram on the right hand side represents $\mathcal{C}_0(t)$. The other two diagrams represent the two distinct rung types that contribute to the ladder sum, see Figure~\ref{laddersum}.}
  \label{ladsum}
\end{figure}

We are now ready to evaluate the ladder sum for $\mathcal{C}$. We start by writing down a Bethe-Saltpeter equation for $\mathcal{C}(t,\x)$ (shown in Figure~\ref{ladsum}). The first diagram on the right (without any rungs) is just given by,
\beq
\mathcal{C}_0(t) = \frac{1}{N} \int d^2\x~ [\CG_R(\x,t)]^2.
\eeq
Chaotic growth comes from the effects of interactions which are captured by the other two diagrams in Figure~\ref{ladsum} which correspond to the two types of rungs discussed above.

We first review some features associated with the type-I rung diagram from the top row in Figure~\ref{ladsum}:
\beq \label{onerung}
\C_{\tn{one-rung}}(\nu) = \frac{1}{N}\int\frac{d\omega}{2\pi}\int_{\p}~\int\frac{d\omega'}{2\pi}\int_{\p'}~\CG_R(\nu-\omega,-\p)~\CG_R(\omega,\p)\nonumber\\~\CG_{W,\lambda}(\omega'-\omega,\p'-\p)~\CG_R(\nu-\omega',-\p')~\CG_R(\omega',\p').
\eeq
The overall prefactor is positive since the two factors of $i$ from the vertex cancel the overall minus sign in the definition and the four factors of $i$ associated with the retarded correlators give unity. The product of the retarded propagators, $\CG_R(\nu-\omega,-\p)~\CG_R(\omega,\p)$, starting with their free-particle forms, is given by
\beq
&&\CG_R(\nu-\omega,-\p)~\CG_R(\omega,\p) =\nonumber \\
&&\frac{1}{4\epsilon_{\p}^2}\bigg(\frac{1}{\nu - \omega - \epsilon_{\p} + i0^+} - \frac{1}{\nu - \omega + \epsilon_{\p} + i0^+} \bigg)~\bigg(\frac{1}{\omega - \epsilon_{\p} + i0^+} - \frac{1}{\omega + \epsilon_{\p} + i0^+} \bigg).\nonumber\\
\eeq
Evaluating the $\omega$ integral using the method of residues gives terms proportional to: $(\nu-2\epsilon_{\p})^{-1},~\nu^{-1},~(\nu+2\epsilon_{\p})^{-1}$.

In order to extract the leading contribution to the long-time form of $\C(t)$, we focus on the most singular $\nu^{-1}$ term. Then we can approximate,
\beq
\CG_R(\nu-\omega,-\p)~\CG_R(\omega,\p) \rightarrow \frac{\pi i}{2\epsilon^2_{\p}} \bigg[\frac{\delta(\omega-\epsilon_{\p})}{\nu + i0^+} + \frac{\delta(\omega + \epsilon_{\p})}{\nu + i0^+} \bigg].
\eeq
The overall positive sign arises from performing the $\omega$ integral by closing it with a clockwise contour in the lower half-plane combined with an explicit minus sign in the residue. At this point, restoring the lifetime from the self-energy correction, we obtain,
\beq
\CG_R(\nu-\omega,-\p)~\CG_R(\omega,\p) \rightarrow \frac{\pi i}{2\epsilon^2_{\p}} \bigg[\frac{\delta(\omega - \epsilon_{\p})}{\nu + 2i\Gamma_{\p}} + \frac{\delta(\omega + \epsilon_{\p})}{\nu + 2i\Gamma_{\p}} \bigg].
\eeq
This combination can be recast as
\beq
\CG_R(\nu-\omega,-\p)~\CG_R(\omega,\p) \rightarrow \frac{\pi}{\epsilon_{\p}} \frac{\delta(\omega^2 - \epsilon_{\p}^2)}{-i \nu + 2 \Gamma_{\p}}.
\eeq

Let us now proceed by summing the ladder series on the first line of Figure~\ref{ladsum} to all orders (i.e. ignoring the diagrams on the second line). We start by defining a function,
\beq
\C(\nu) = \frac{1}{N}\int \frac{d\omega}{2\pi}\int_{\p} ~f(\nu;\omega,\p).
\eeq
Then, the Bethe-Saltpeter equation becomes,
\beq
f(\nu;\omega,\p) = \CG_R(\nu-\omega,-\p)~\CG_R(\omega,\p)\bigg[1 + \frac{1}{N}\int\frac{d\omega'}{2\pi} \int_{\l} \CG_{W,\lambda}(\omega'-\omega,\l-\p)~f(\nu;\omega',\l)  \bigg].\nonumber\\
\eeq

To estimate the behavior at large time, we drop the homogenous term; assuming that $f(t)$ grows exponentially, the ladder sum should remain invariant under the addition of an extra rung to the series. Therefore, substituting the form for the propagators,
\beq
(-i\nu + 2\Gamma_{\p})~f(\nu;\omega,\p) \approx \frac{1}{N}\frac{\pi}{\epsilon_{\p}}~\delta(\omega^2-\epsilon^2_{\p})~\int\frac{d\omega'}{2\pi} \int_{\l}~ \CG_{W,\lambda}(\omega'-\omega,\l-\p)~f(\nu;\omega',\l).
\eeq
It is natural to postulate the following on-shell form of the function $f$:
\beq
f(\nu;\omega,\p) = f(\nu,\p)~\delta(\omega^2-\epsilon^2_{\p}).
\label{onshell}
\eeq
The integral equation for the ladder sum then becomes,
\beq
(-i\nu + 2\Gamma_{\p})~f(\nu,\p) &\approx& \frac{1}{N} \int_{\l}   \Ru_1(\l,\p)~f(\nu,\l),\\
\Ru_1(\l,\p) &=& R_{1,+}(\l,\p) + R_{1,-}(\l,\p),\\
R_{1,\pm}(\l,\p) &=& \frac{\CG_{W,\lambda}(\pm\epsilon_{\l} - \epsilon_{\p},\l-\p)}{4\epsilon_{\p}\epsilon_{\l}}.
\label{rung}
\eeq
Finally, plugging in the explicit form of $\Gamma_{\p}$ from Eq.~\eqref{gammaq}, we obtain the following integral equation,
\beq
-i\nu~f(\nu,\p) &\approx& \frac{1}{N} \int_{\l} \Ru_1(\l,\p)  \bigg[ f(\nu,\l) - \frac{\sinh(\beta\epsilon_{\p}/2)}{\sinh(\beta\epsilon_{\l}/2)} f(\nu,\p) \bigg].
\label{rung2}
\eeq

Having understood the general structure of the ladder sum, let us now include the type-II rung. Just as in the previous computation, we begin by considering the one-block diagram:
\beq
\C_{\tn{one-block}}(\nu) = \frac{1}{N}\int\frac{d\omega}{2\pi}\int_{\p}~\int\frac{d\omega''}{2\pi}\int_{\p''}~\int\frac{d\omega'}{2\pi}\int_{\p'} ~\CG_R(\nu-\omega,-\p)~\CG_R(\omega,\p)\nonumber\\~\CG_{W}(\omega''-\omega,\p''-\p)~\CG_{R,\lambda}(\nu-\omega'',-\p'')~\CG_{R,\lambda}(\omega'',\p'')~\CG_W(\omega'-\omega'',\p'-\p'')\nonumber\\
~\CG_R(\nu-\omega',-\p')~\CG_R(\omega',\p').
\eeq
Note that the overall sign is again positive. It is easiest to proceed by defining,
\beq
\CG_{\tn{eff}}(\omega',\omega,\p',\p) = \int\frac{d\omega''}{2\pi}\int_{\p''}\CG_{W}(\omega''-\omega,\p''-\p)~\CG_W(\omega'-\omega'',\p'-\p'')\nonumber\\
~\CG_{R,\lambda}(\nu-\omega'',-\p'')~\CG_{R,\lambda}(\omega'',\p''),
\label{Geff}
\eeq
so that the one-block diagram becomes,
\beq
\C_{\tn{one-block}}(\nu) = \frac{1}{N}\int\frac{d\omega}{2\pi}\int_{\p}~\int\frac{d\omega'}{2\pi}\int_{\p'} ~\CG_R(\nu-\omega,-\p)~\CG_R(\omega,\p)\nonumber\\
~\CG_{\tn{eff}}(\omega',\omega,\p',\p)~\CG_R(\nu-\omega',-\p')~\CG_R(\omega',\p').
\eeq
The one-block and one-rung diagrams can be grouped into one contribution by shifting $\CG_{W,\lambda}(\omega'-\omega,\p'-\p)\rightarrow \CG_{W,\lambda}(\omega'-\omega,\p'-\p) + \CG_{\tn{eff}}(\omega',\omega,\p',\p)$; therefore we can simply proceed by setting up the ladder sum as before after shifting the Wightman function as above. Note that we do not assume $\CG_{\tn{eff}}$ to be a function of only the relative frequency or momentum.

The next step is to evaluate $\CG_{\text{eff}}$. Recall Eq.~\eqref{wightspec} which is $\CG_W(\omega,\k) = {\cal{Q}}(\omega)~A(\omega,\k)$, where ${\cal{Q}}(\omega) = [2\sinh(\beta\omega/2)]^{-1}$. In this formula it is sufficient to use the bare $\varphi$ spectral function since the rung already contains an explicit factor of $1/N$. Inserting the spectral function gives
\beq
\CG_{W}(\omega''-\omega,\p''-\p)~\CG_W(\omega'-\omega'',\p'-\p'') = {\cal{Q}}(\omega''-\omega)~{\cal{Q}}(\omega'-\omega'')~\frac{\pi^2}{\epsilon_{\p''-\p}~\epsilon_{\p'-\p''}} \nonumber\\
\bigg[\delta(\omega'' - \omega - \epsilon_{\p''-\p}) - \delta(\omega'' - \omega + \epsilon_{\p''-\p})\bigg]\nonumber\\
\bigg[\delta(\omega' - \omega'' - \epsilon_{\p'-\p''}) - \delta(\omega' - \omega'' + \epsilon_{\p'-\p''})\bigg].
\eeq
Denote the relative momentum, $\p'-\p = \bar\p$, and the centre-of-mass momentum, $\P = (\p+\p')/2$. Also define $\bar\omega = \omega'-\omega$ and shift $\p''\rightarrow\p''+\P$. We can now use one of the delta functions above to integrate over $\omega''$ in Eq.~\eqref{Geff}. This leads to,
\beq
\CG_{\tn{eff}}(\omega',\omega,\bar\p,\P) = \frac{1}{2N}\int_{\p''} \frac{\pi}{\epsilon_{\frac{\bar\p}{2}+\p''}~\epsilon_{\frac{\bar\p}{2}-\p''}}  \bigg({\cal{Q}}(\epsilon_{\p''+\frac{\bar\p}{2}})~{\cal{Q}}(\bar\omega - \epsilon_{\p''+\frac{\bar\p}{2}})\nonumber\\
\CG_{R,\lambda}(\nu-\omega-\epsilon_{\p''+\frac{\bar\p}{2}},-\p''-\P)~\CG_{R,\lambda}(\omega+\epsilon_{\p''+\frac{\bar\p}{2}},\p''+\P)\nonumber\\
\bigg[\delta(\bar\omega - \epsilon_{\frac{\bar\p}{2}+\p''} - \epsilon_{\frac{\bar\p}{2}-\p''}) - \delta(\bar\omega - \epsilon_{\frac{\bar\p}{2}+\p''} + \epsilon_{\frac{\bar\p}{2}-\p''})\bigg]~\nonumber\\
- {\cal{Q}}(-\epsilon_{\p''+\frac{\bar\p}{2}})~{\cal{Q}}(\bar\omega + \epsilon_{\p''+\frac{\bar\p}{2}})\nonumber\\
\CG_{R,\lambda}(\nu-\omega+\epsilon_{\p''+\frac{\bar\p}{2}},-\p''-\P)~\CG_{R,\lambda}(\omega-\epsilon_{\p''+\frac{\bar\p}{2}},\p''+\P)\nonumber\\
\bigg[\delta(\bar\omega + \epsilon_{\frac{\bar\p}{2}+\p''} - \epsilon_{\frac{\bar\p}{2}-\p''}) - \delta(\bar\omega + \epsilon_{\frac{\bar\p}{2}+\p''} + \epsilon_{\frac{\bar\p}{2}-\p''})\bigg]\bigg).
\eeq

The next step is to set $\nu=0$ in the above integrand; one can ignore the $\nu$ dependence to leading order since $\nu$ is the external frequency and it will ultimately be of order $1/N$. Denoting $\epsilon_{\frac{\bar\p}{2} \pm \p''} = \epsilon_\pm$, the above can be expressed as,
\beq
\CG_{\tn{eff}}(\omega',\omega,\bar\p,\P) = \frac{1}{2N}\int_{\p''} \frac{\pi}{\epsilon_{+}~\epsilon_{-} } \bigg({\cal{Q}}(\epsilon_{+})~{\cal{Q}}(\bar\omega - \epsilon_{+})\nonumber\\
\CG_{R,\lambda}(-\omega-\epsilon_{+},-\p''-\P)~\CG_{R,\lambda}(\omega+\epsilon_{+},\p''+\P)\nonumber\\
\bigg[\delta(\bar\omega - \epsilon_{+} - \epsilon_{-}) - \delta(\bar\omega - \epsilon_{+} + \epsilon_{-})\bigg]~\nonumber\\
- {\cal{Q}}(-\epsilon_{+})~{\cal{Q}}(\bar\omega + \epsilon_{+})\nonumber\\
\CG_{R,\lambda}(-\omega+\epsilon_{+},-\p''-\P)~\CG_{R,\lambda}(\omega-\epsilon_{+},\p''+\P)\nonumber\\
\bigg[\delta(\bar\omega + \epsilon_{+} - \epsilon_{-}) - \delta(\bar\omega + \epsilon_{+} + \epsilon_{-})\bigg]\bigg).
\label{Geff2}
\eeq

The delta function constraints in the above equation are of the form,
\beq
\bar\omega = s_1 \epsilon_+ + s_2 \epsilon_-, ~~(s_1, s_2 = \pm 1).
\eeq
Squaring both sides, we have,
\beq
\bar\omega^2 - 2 \bar\omega s_1 \epsilon_+ + \epsilon_+^2 = \epsilon_-^2,
\eeq
leading to
\beq
|\p''| = |\p''|_0 = \frac{\bar\omega}{2}\sqrt{\frac{\bar\omega^2 - |\bar\p|^2 - 4\mu^2}{\bar\omega^2 - |\bar\p|^2\cos^2\theta}},
\eeq
where we have chosen the physical positive root and $\theta$ is the angle between $\p''$ and $\overline\p$. We can carry out the integral over the magnitude of $\p''$ using the delta function. The final integral over the orientation, i.e. the angle $\theta$ between $\p''$ and $\bar\p$ must be done numerically. The use of the delta function and the subsequent angular integral parallel the computation of the bubble in Appendix \ref{oneloop}.

The fully re-summed ladder with both rung types is obtained from the solution of
\beq
(-i\nu + 2\Gamma_{\p})~f(\nu;\omega,\p) \approx \frac{1}{N}\frac{\pi}{\epsilon_{\p}}~\delta(\omega^2-\epsilon^2_{\p}) \nonumber \\
\times ~\int\frac{d\omega'}{2\pi} \int_{\l}~\bigg[  \CG_{W,\lambda}(\omega'-\omega,\l-\p) + \CG_{\tn{eff}}\bigg(\omega',\omega,\l-\p,\frac{\l+\p}{2}\bigg) \bigg]~f(\nu;\omega',\l).
\label{resummed}
\eeq
Postulating the on-shell form in Eq.~\eqref{onshell} for $f$, the integral equation for the ladder sum becomes,
\beq
(-i\nu + 2\Gamma_{\p})~f(\nu,\p) &\approx& \frac{1}{N} \int_{\l} \left\{\Ru_1(\l,\p) + \Ru_2(\l,\p)\right\}~f(\nu,\l),\\
\Ru_{2}(\l,\p) &=& R_{2,+}(\l,\p) + R_{2,-}(\l,\p),\\
R_{2,\pm}(\l,\p) &=& \frac{1}{4\epsilon_{\p}\epsilon_{\l}}\CG_{\tn{eff}}\bigg(\pm\epsilon_{\l},\epsilon_{\p},\l-\p,\frac{\l+\p}{2}\bigg). \nonumber\\
\label{rung}
\eeq
Finally, together with the explicit form of $\Gamma_{\p}$ from Eq.~\eqref{gammaq}, we obtain the following integral equation,
\beq
-i\nu~f(\nu,\p) &=& \frac{1}{N} \int_{\l}  \bigg[ \Ru_1(\l,\p) + \Ru_2(\l,\p) - 2 N \Gamma_{\vec{p}} (2\pi)^2 \delta^{(2)}(\vec{\ell}-\vec{p}) \bigg] f(\nu,\vec{\ell}),
\label{rung3}
\eeq
where we have factored out the explicit $1/N$ dependence of $\Gamma_{\vec{p}}$ above.

Some comments are in order. First, note that both $\Ru_1$ and $\Ru_2$ are positive and tend to cause chaos to grow while $\Gamma_{\vec{p}}$ suppresses the growth. Besides the explicit momenta, the only energy scale appearing in $\Ru_1$ and $\Ru_2$ is the thermal mass $\mu( \propto T)$ and the temperature itself, so in fact the only scale is the temperature. By rescaling all momenta and frequencies by a factor of $T$, it immediately follows that the self consistent ladder equation is
\beq
- i \nu f = \frac{T}{N} \widehat{\mathcal{M}} f
\eeq
where $\widehat{\mathcal{M}}$ is a scaled integral kernel in which the temperature has been set to one. Hence without any further analysis it follows that if a growing mode exists, then its growth exponent is proportional to $T/N$.

\section{Chaos exponent}
\label{LE}

By substituting $-i \nu \rightarrow \partial_t$, we now view Eq.~\eqref{rung3} as a matrix differential equation
\beq
\partial_t f = \frac{T}{N} \widehat{\mathcal{M}} f,
\eeq
where $\widehat{\cal{M}}$ represents the contributions from both rungs and the damping. By further rescaling the frequency, we obtain a dimensionless integral equation which is suitable for numerical analysis. The growth of $f$ and hence $\cal{C}$ will then be determined by the largest positive eigenvalue of $\widehat{\cal{M}}$. All of the individual ingredients in $\widehat{\cal{M}}$, i.e. the rung functions, can be computed numerically starting from the polarization bubble.

We now summarize the numerical procedure used to solve the eigenvalue equation
\beq
\lambda_i f_i = \frac{T}{N} \widehat{\cal{M}}f_i.
\eeq
Our main assumption is that the eigenvector $f_1$ associated with the {\it largest} eigenvalue $\lambda_1 \equiv \lambda_L$ is rotationally invariant, $f_1(\p) = f_1(|\p|)$. Then the eigenvalue equation can be simplified by projecting both sides onto states with angular momentum $l=0$, i.e. `$s$-wave' states. The resulting eigenvalue equation involves an integral transform of $f$ which is one dimensional, involving only the magnitude of the momentum. The assumption of rotational invariance essentially permits a more precise numerical computation of the spectrum of $\cal{M}$ within the rotationally invariant subspace.

The strategy is then three-fold. First, we compute the spectral function $A_\lambda$ numerically on a fine grid in $(\omega,|\q|)$ space. Second, we use those data to construct the rung functions and to perform an approximate angular integral in the eigenvalue equation. Third, we solve the eigenvalue equation on a discretized grid in the radial momentum. Since the rung functions effectively vanish when the norms of their arguments greatly exceed $T$, we use a simple linearly spaced grid with a hard momentum cutoff of order $10~ T$. As discussed above, the temperature is the only scale in the problem, so the entire numerical computation is set up in terms of the scaled variables $\mathfrak{q} = q/T$ and $\mathfrak{w} = \omega/T$.

\begin{figure}
  \centering
  \includegraphics[width=.8\textwidth]{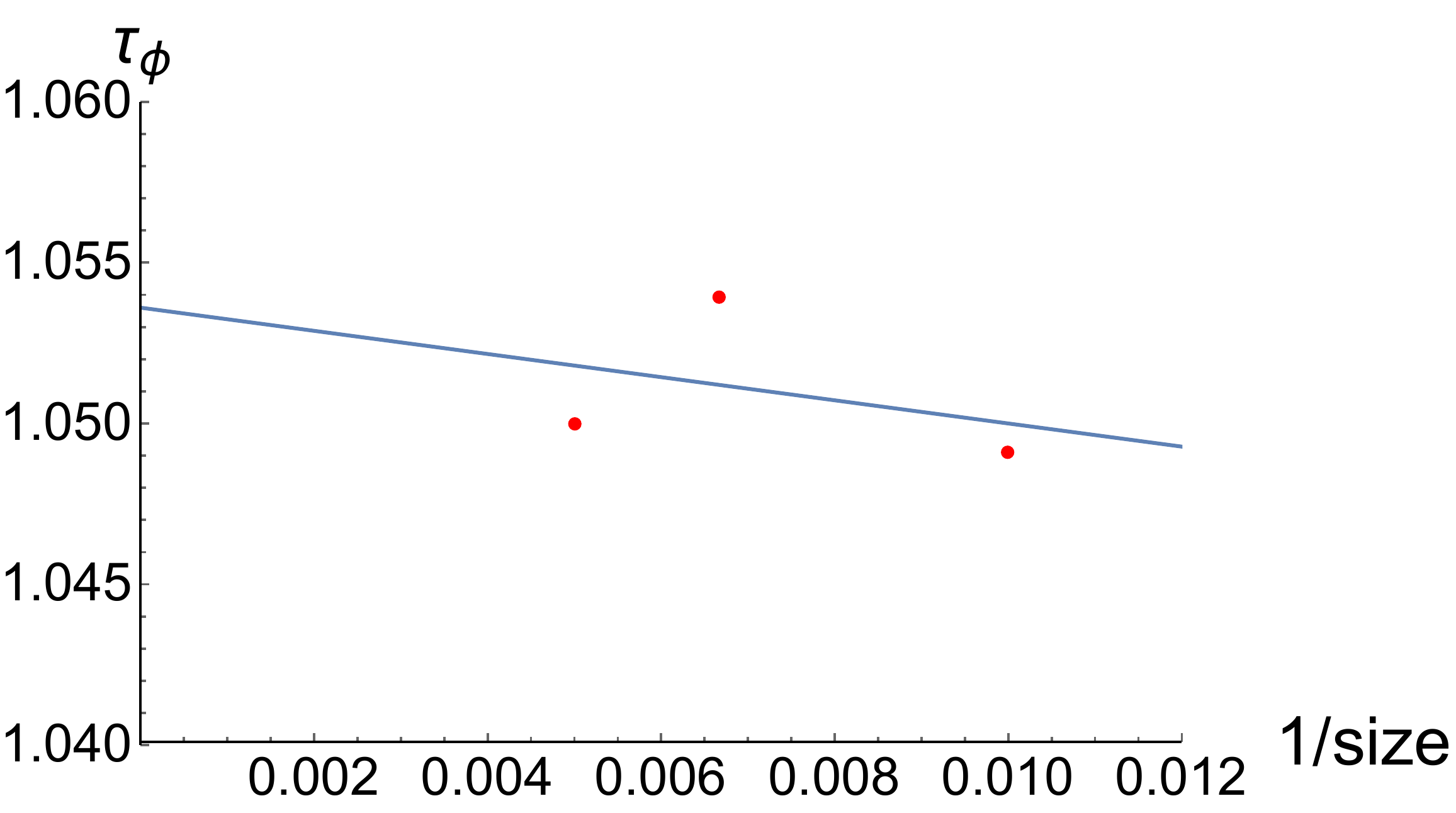}
  \caption{The dephasing rate (in units of $T/N$) versus inverse grid size computed using the rotationally invariant scheme (red points). Also shown is a linear fit to the data. The extrapolated value is $1/\tau_\varphi \approx 1.05 ~T/N$ which is approximately $10$\% off from the high precision value. There is some uncertainty in the extrapolated value as a result of the scatter associated with the data points. However, even in the worst case scenario, the uncertainty is at the level of approximately only a few percent.}
  \label{fig:Gk0}
\end{figure}

\begin{figure}
  \centering
  \includegraphics[width=.85\textwidth]{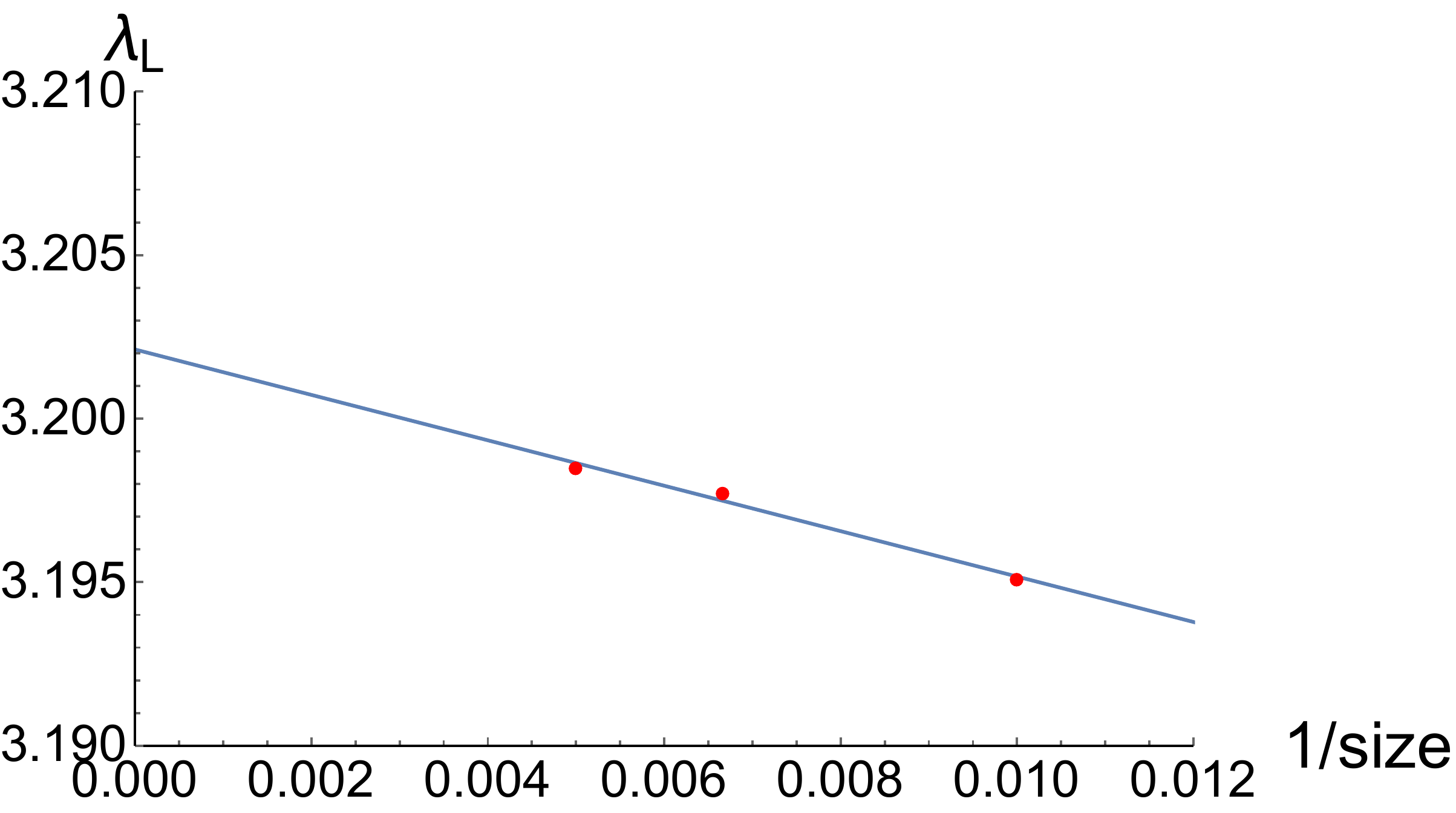}
  \caption{The chaos exponent (in units of $T/N$) versus inverse grid size computed using the rotationally invariant scheme (red points). Also shown is a linear fit to the data. The extrapolated value is $\lambda_L \approx 3.2~T/N$.}
  \label{fig:Lk0}
\end{figure}

The data for $1/\tau_\varphi$ and $\lambda_L$ as a function of grid size are shown in Figures \ref{fig:Gk0} and \ref{fig:Lk0}. We roughly estimate, based on a comparison with the high precision value of $1/\tau_\varphi$, that our results are accurate to roughly $10$\%. However, we caution the reader that $\tau_\varphi$ is not sensitive to the type-II rung, so this is only a partial benchmark.

We emphasize that a positive growth exponent proportional to $T/N$ is a robust prediction of our calculation. To better explain the limitations of our calculation and the sources of error, we provide additional details of the calculation in Appendix \ref{sec:numerics}.

\section{Butterfly velocity}
\label{VB}

In this section we go back to studying $\mathcal{C}(t,\x)$ when $\x$ is not integrated over, which is given by
\beq
\mathcal{C}(t,\x) = - \frac{1}{N^2} \sum_{a,b} \text{Tr}\left\{\sqrt{\rho} ~[\varphi_a(\x,t),\varphi_b] \sqrt{\rho} ~[\varphi_a(\x,t), \varphi_b] \right\}.
\label{ct}
\eeq
This amounts to evaluating the ladder-sum as earlier, but at a finite external momentum. For the purpose of illustration, we start again by considering only the type-I rung (first line of Figure~\ref{ladsum}); the generalization due to the type-II rung is straightforward.

At finite external momentum, the product of the $\varphi$ propagators is
\beq
&&\CG_R(\nu-\omega,\k-\p)~\CG_R(\omega,\p) =\nonumber \\
&&\frac{1}{4\epsilon_{\p} \epsilon_{\k-\p}}\bigg(\frac{1}{\nu - \omega - \epsilon_{\k-\p} + i0^+} - \frac{1}{\nu - \omega + \epsilon_{\k-\p} + i0^+} \bigg)~\bigg(\frac{1}{\omega - \epsilon_{\p} + i0^+} - \frac{1}{\omega + \epsilon_{\p} + i0^+} \bigg).\nonumber\\
\eeq
Evaluating the integral over $\omega$ by residue again gives rise to a variety of terms; the long time behavior is well approximated by retaining just two terms,
\beq
\CG_R(\nu-\omega,\k-\p)~\CG_R(\omega,\p) \rightarrow \frac{\pi i}{2\epsilon_{\p}\epsilon_{\k-\p}} \bigg[\frac{\delta(\omega-\epsilon_{\p})}{\nu - (\epsilon_{\p} - \epsilon_{\k-\p}) + i0^+} + \frac{\delta(\omega + \epsilon_{\p})}{\nu + \epsilon_{\p} - \epsilon_{\k-\p} + i0^+} \bigg].\nonumber\\
\eeq
The overall sign is again positive due to the clockwise contour and the explicit minus sign in the residue.

These two terms generalize the $1/\nu$ terms from the $\k=0$ calculation; they are the most singular terms at small $\nu$ and $\k$ and hence are expected to lead to the fastest growth. Restoring the self-energy correction but ignoring the correction due to a finite $\k$ on the damping, i.e. $\Gamma_{\k-\p} \approx \Gamma_{\p}$, we have,
\beq
\CG_R(\nu-\omega,\k-\p)~\CG_R(\omega,\p) \rightarrow \frac{\pi i}{2\epsilon_{\p}\epsilon_{\k-\p}} \bigg[\frac{\delta(\omega-\epsilon_{\p})}{\nu - (\epsilon_{\p} - \epsilon_{\k-\p}) + 2i\Gamma_{\p}} + \frac{\delta(\omega + \epsilon_{\p})}{\nu + \epsilon_{\p} - \epsilon_{\k-\p} + 2i\Gamma_{\p}} \bigg].\nonumber
\eeq

Let us now proceed to evaluate the ladder-sum. We first define (as was done previously),
\beq
\mathcal{C}(\nu, \k) = \frac{1}{N} \int \frac{d\omega}{2\pi} \int_{\p} f(\nu,\k; \omega,\p),
\eeq
such that the Bethe-Saltpeter equation for the first line in Figure~\ref{ladsum} for $f$ becomes,
\beq
 f(\nu,\k; \omega,\p) = \CG_R(\nu-\omega,\k-\p)~\CG_R(\omega,\p) \bigg[1 + \frac{1}{N} \int_{\p'} \CG_{W,\lambda}(\omega'-\omega,\p'-\p) ~f(\nu,\k;\omega',\p')  \bigg]. \nonumber \\
 \label{laddersp}
\eeq
We ignore the homogenous term as before and make the following ansatz for $f(\nu,\k; \omega,\p)$:
\beq
 f(\nu,\k; \omega,\p) = \frac{f_+(\nu,\k;\p)}{2\epsilon_{\p}} \delta(\omega-\epsilon_{\p}) + \frac{f_-(\nu,\k;\p)}{2\epsilon_{\p}} \delta(\omega+\epsilon_{\p}).
\eeq
Roughly speaking, the quantities $f_+$ and $f_-$ are analogous to the density of particles and holes in a Boltzmann-like equation, to be derived shortly. The ``on-shell" condition now accounts for propagation with two opposite velocities, as will become clear soon. Note that the above reduces to the previous ansatz in the limit of $\k\rightarrow0$, where $f_+(\nu,\vec{0},\p) = f_-(\nu,\vec{0},\p) = f(\nu,\p)$.

The ladder sum in Eq.~\eqref{laddersp} becomes,
\beq
(-i\nu + i\delta\epsilon_{\k} + 2\Gamma_{\p}) f_+(\nu,\k;\p) = \frac{1}{N}\int_{\p'}\bigg[\Ru_{1,+}(\p',\p) f_+(\nu,\k;\p') + \Ru_{1,-}(\p',\p) f_-(\nu,\k;\p') \bigg], \nonumber\\ \\
(-i\nu - i\delta\epsilon_{\k} + 2\Gamma_{\p}) f_-(\nu,\k;\p) = \frac{1}{N}\int_{\p'}\bigg[\Ru_{1,-}(\p',\p) f_+(\nu,\k;\p') + \Ru_{1,+}(\p',\p) f_-(\nu,\k;\p') \bigg],\nonumber\\
\label{vblad}
\eeq
where $\Ru_{1,\pm}(\p',\p)$ are given by Eq.~\eqref{rung}. We have made the following simplifications in arriving at the above equations: $\epsilon_{\p'}\epsilon_{\k-\p} \approx \epsilon_{\p}\epsilon_{\p'}$ and $\epsilon_{\p}-\epsilon_{\k-\p} \equiv \delta\epsilon_{\k} \approx \k\cdot v_{\p}$, where $v_{\p}=\nabla_{\p}\epsilon_{\p}$ is the group velocity.

The analogous equations with the second rung included are obtained by replacing $\Ru_{1,\pm}$ by $\Ru_{1,\pm} + \Ru_{2,\pm}$. From now on, the second rung will be included in the calculation. It is convenient to define $\Ru_{\pm} = \Ru_{1,\pm} + \Ru_{2,\pm}$.

The above two equations can be combined into a compact expression by introducing the vector $\Psi^T(\nu,\k;\p) = (f_+(\nu,\k;\p)~~f_-(\nu,\k;\p))$,
and the matrix,
\beq
\widehat\Ru(\p',\p) = \left(\begin{array}{cc}
\Ru_+(\p',\p) & \Ru_-(\p',\p) \\
\Ru_-(\p',\p) & \Ru_+(\p',\p)  \end{array} \right).
\eeq
The equation for $\Psi$ then takes a form that is similar to a kinetic theory Boltzmann equation,
\beq
(-i\nu + 2\Gamma_{\p}) \hat\sigma_0 \Psi(\nu,\k;\p) + i\delta\epsilon_{\k} \hat\sigma_z \Psi(\nu,\k;\p) = \frac{1}{N}\int_{\p'} \widehat\Ru(\p',\p) \Psi(\nu,\k;\p'),
\eeq
Finally, recall that $\Gamma_{\p} ~(\sim 1/N)$ can itself be expressed in terms of $\Ru_1$.

The matrix structure can be further simplified with the ansatz that the ``particle" density at momentum $\vec{p}$ is equal to the ``hole" density at momentum $-\vec{p}$, i.e. $f_+(\nu,\k;-\p) = f_-(\nu,\k;\p)$. One can then decouple the equations for $f_\pm$ in Eq.~\eqref{vblad} as,
\beq
(-i\nu + i\delta\epsilon_{\k} + 2\Gamma_{\p}) f_+(\nu,\k;\p) &=& \frac{1}{N}\int_{\p'}\bigg[\Ru_{+}(\p',\p) + \Ru_{-}(-\p',\p) \bigg]f_+(\nu,\k;\p') , \nonumber\\ \\
(-i\nu - i\delta\epsilon_{\k} + 2\Gamma_{\p}) f_-(\nu,\k;\p) &=& \frac{1}{N}\int_{\p'}\bigg[\Ru_{-}(-\p',\p)  + \Ru_{+}(\p',\p) \bigg]f_-(\nu,\k;\p'). \nonumber\\
\label{decoupled}
\eeq
The final form of the finite $\k$ equation, let us say for $f_+$, is thus
\beq
\partial_t f_+(\p) + i \k \cdot v_{\p} ~f_+(\p) = \frac{1}{N} \int_{\p'} \widehat{\mathcal{K}}(\p',\p) ~f_+(\p'),
\label{decoup_fin}
\eeq
where we use the shorthand notation $f_+(\p)\equiv f_+(t,\k;\p)$ and the expression for $\widehat{\mathcal{K}}(\p',\p)$ can be read off from Eq.~\eqref{decoupled},
\beq
\widehat{\mathcal{K}}(\p',\p) = \Ru_{+}(\p',\p) +  \Ru_{-}(-\p',\p) -  2 N \Gamma_{\vec{p}} (2\pi)^2 \delta^{(2)}(\vec{p}'-\vec{p}).
\eeq
Note that we have factored out the $1/N$ dependence in $\Gamma_{\vec{p}}$ above.

We analyze Eq. \eqref{decoup_fin} by thinking of the $\k$ term as a perturbation. To do this consistently, we should first render $\k$ dimensionless. Let $\lambda_0 \sim T/N$ denote the largest eigenvalue of $\widehat{\mathcal{K}}$ times the temperature at $\k=0$. Then $1/\lambda_0$ defines a length-scale (recall that $c=1$) associated with the inelastic physics of growing chaos. Just like the dephasing-time or length-scale, $1/\Gamma_0 = \tau_\varphi$, the length-scale $1/\lambda_0$ is proportional to $N/T$ and hence grows at large $N$. For simplicity we thus introduce the dimensionless variable
\beq
{\vec u} = \frac{N \k}{T}
\eeq
and consider perturbation theory in small ${\vec u}$.

Treating ${\vec u}$ as small amounts to considering distances that are long compared to $1/\lambda_0$ and operators that are smeared on the scale of $1/\lambda_0$. This is a physical requirement because there is an important order of limits issue. For distances smaller than $1/\lambda_0$, the physics is that of free particles, so our analysis of the chaotic growth of operators only applies for times and distances longer than $1/\lambda_0$.

We now consider the leading eigenvalue $\lambda_L({\vec u})$ for small ${\vec u}$. The leading result is of course just $\lambda_0$ and the first order correction vanishes because of the spherical symmetry of the leading eigenvector at $\k=0$. Equivalently, we are assuming that the largest eigenvalue is a function of only $u = |{\vec u}|$. The first correction occurs at second order in $u$,
\beq \label{lambdaquad}
\lambda_L({\vec u}) \approx \lambda_0 - \lambda_2 u^2 + ...
\eeq
In this formula, $\lambda_0$ and $\lambda_2$ are positive and proportional to $T/N$.

Returning to $\mathcal{C}(t,{\k})$, we find that it grows at the rate $\lambda_L({\vec u})$. $\mathcal{C}(t,\x)$ is obtained by taking the Fourier transform,
\beq
\mathcal{C}(t,\x) = \int_{\k} e^{i \k \cdot \x}~ \mathcal{C}(t,\k).
\eeq
As discussed in Eq. \ref{poles}, we assume that the physically relevant dependence on $\k$ comes from the exponential growth factor (i.e. is dominated by the saddle-point $\lambda_L(\vec{k})$, and not by the eigenvector). We then estimate
\beq
\mathcal{C}(t,\x) \sim \int_{\k} \exp\left(i \k \cdot \x + \lambda_0 t - \frac{N^2 \lambda_2 }{T^2} |\k|^2 t \right).
\eeq
The result of the $\k$ integral is
\beq
\mathcal{C}(t,\x) \sim \exp\left(\lambda_L t - \frac{|\x|^2}{4 D_L t} \right)
\label{finctx}
\eeq
where
\beq
\lambda_L = \lambda_0
\eeq
and
\beq
D_L = \frac{N^2 \lambda_2}{T^2}.
\eeq
Recall that $\lambda_0, \lambda_2 \sim T/N$, which implies that $D_L \sim N/T$.

Balancing the two terms in the exponent of $\mathcal{C}(t,\x)$ in Eq. \eqref{finctx} gives a ballistic condition
\beq
|\x|^2 = 4 D_L \lambda_L t^2
\eeq
which translates to a butterfly velocity
\beq
v_B &=& \sqrt{4 D_L \lambda_L} = \sqrt{4\tilde\lambda_2\tilde\lambda_0},\\
\tilde\lambda_{0,2} &=& \frac{N}{T} \lambda_{0,2}.
\eeq
Since $\lambda_0$ and $\lambda_2$ both scale like $T/N$, it follows that $v_B$ is independent of $T$ and $N$. This is consistent with $v_B \sim v_0 T^{1-1/z}$ obtained from simple scaling arguments with $[x] = -1$ and $[t] = -z = -1$, where $z$ is the dynamical exponent ($v_0$ is some natural velocity scale in the problem, appropriately dimensionalized) and has been pointed out in earlier holographic studies of quantum chaos \cite{Blake16} \cite{BSLR}. There is a subtle order of limits issue here as discussed above; the result for $v_B$ is only valid on scales longer than $1/\lambda_0$, which diverges as $N \rightarrow \infty$, i.e. as we approach the free particle limit.

We may also take the exponent of $\mathcal{C}(t,\x)$, write $t = |\x|/v_B + (t - |\x|/v_B)$, and expand in $t - |\x|/v_B$ (valid near the wavefront) to yield
\beq
\mathcal{C}(t,\x) \sim \exp\left( 2 \lambda_L [t - |\x|/v_B ] + ...\right).
\eeq

\begin{figure}
  \centering
  \includegraphics[width=.8\textwidth]{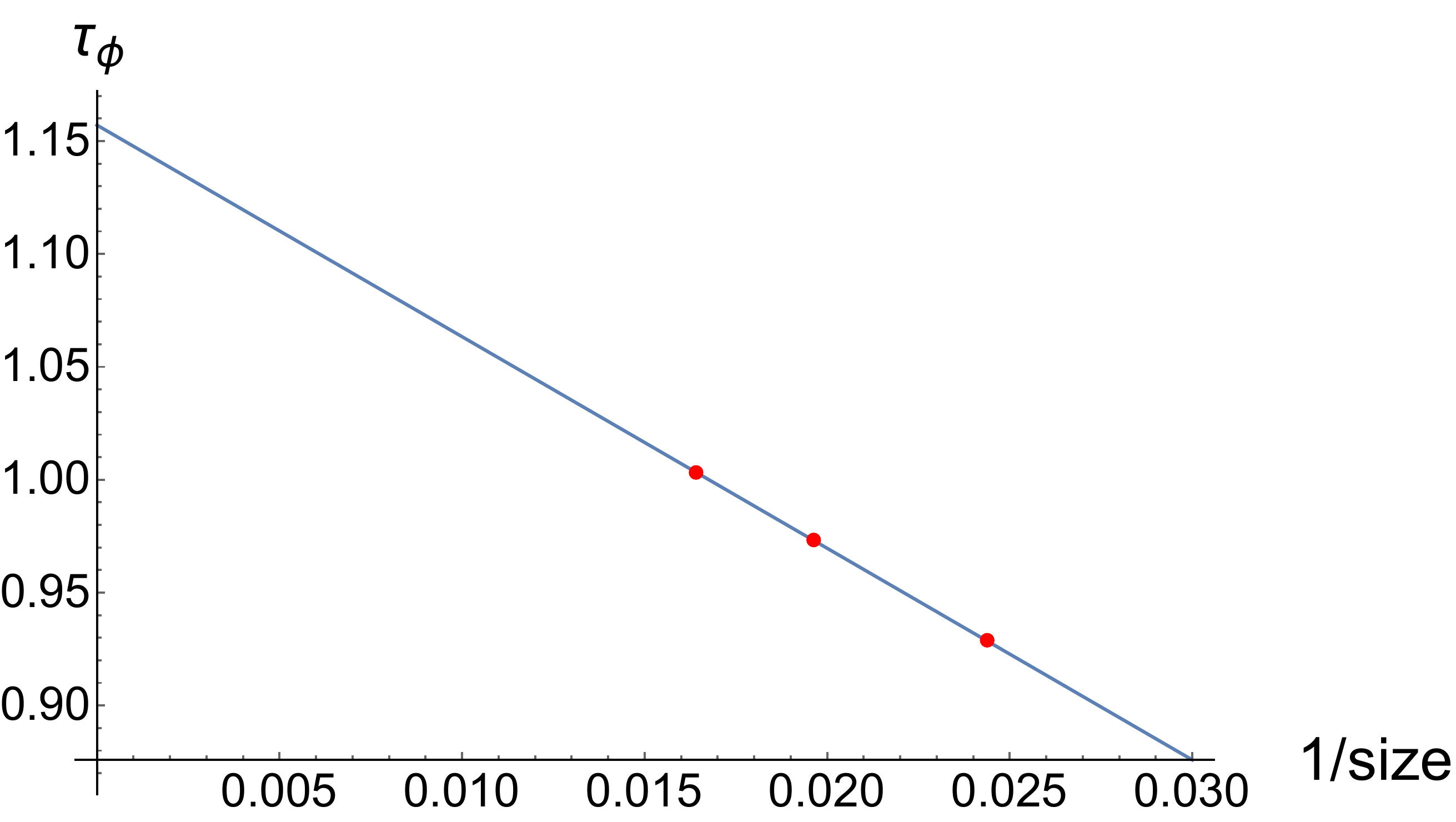}
  \caption{Dephasing rate (in units of $T/N$) versus inverse grid size computed without assuming rotational invariance. Also shown is a linear fit to the data. The extrapolated value is $1/\tau_\varphi \approx 1.16~T/N$ which is approximately $1$\% off from the high precision value.}
  \label{fig:Gk}
\end{figure}
\begin{figure}
  \centering
  \includegraphics[width=.85\textwidth]{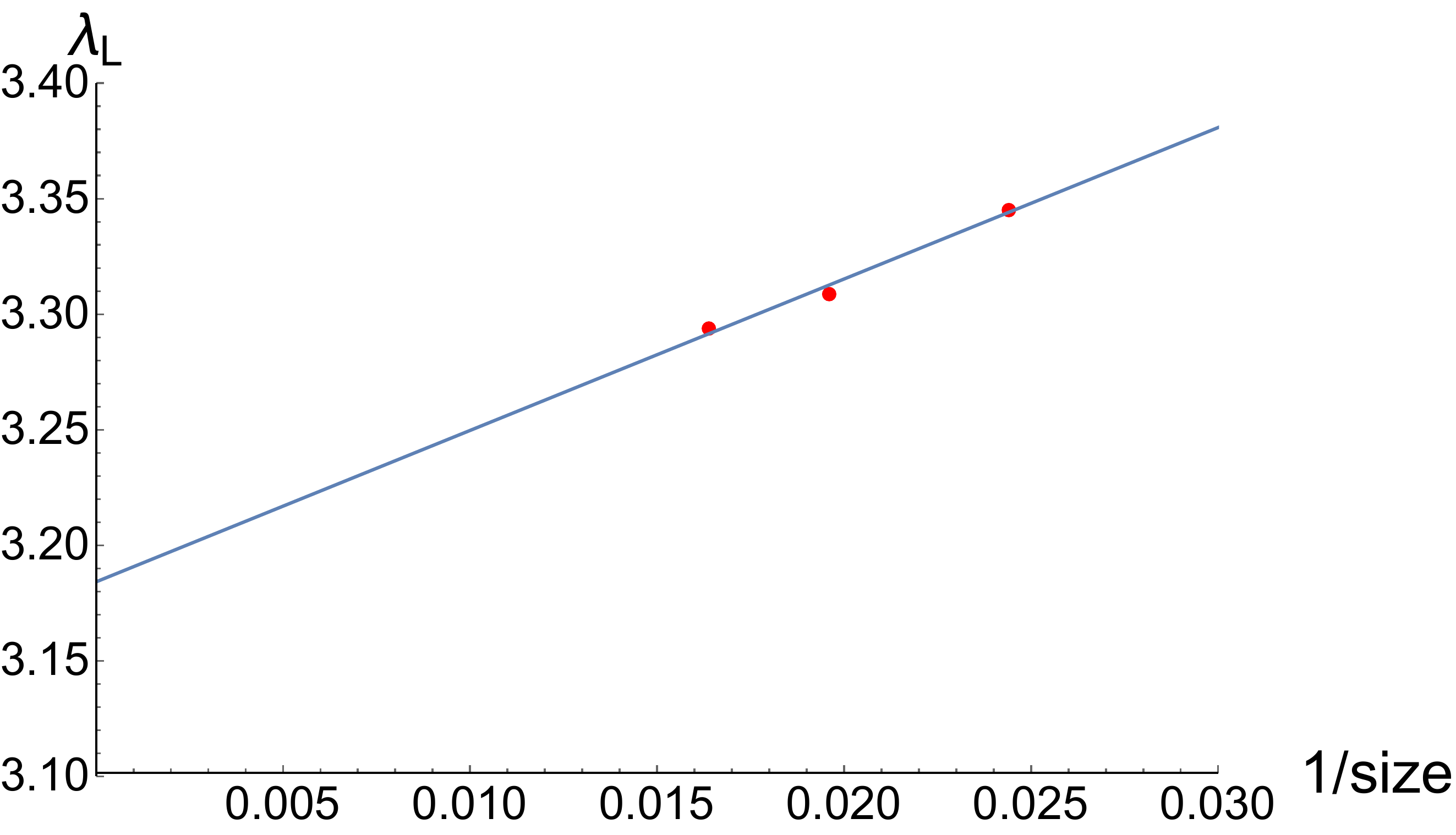}
  \caption{Chaos exponent (in units of $T/N$) versus inverse grid size computed without assuming rotational invariance. Also shown is a linear fit to the data. The extrapolated value is $\lambda_L \approx 3.2~T/N$ which agrees well with the rotationally invariant computation (see Figure \ref{fig:Lk0}).}
  \label{fig:Lk}
\end{figure}
\begin{figure}
  \centering
  \includegraphics[width=.8\textwidth]{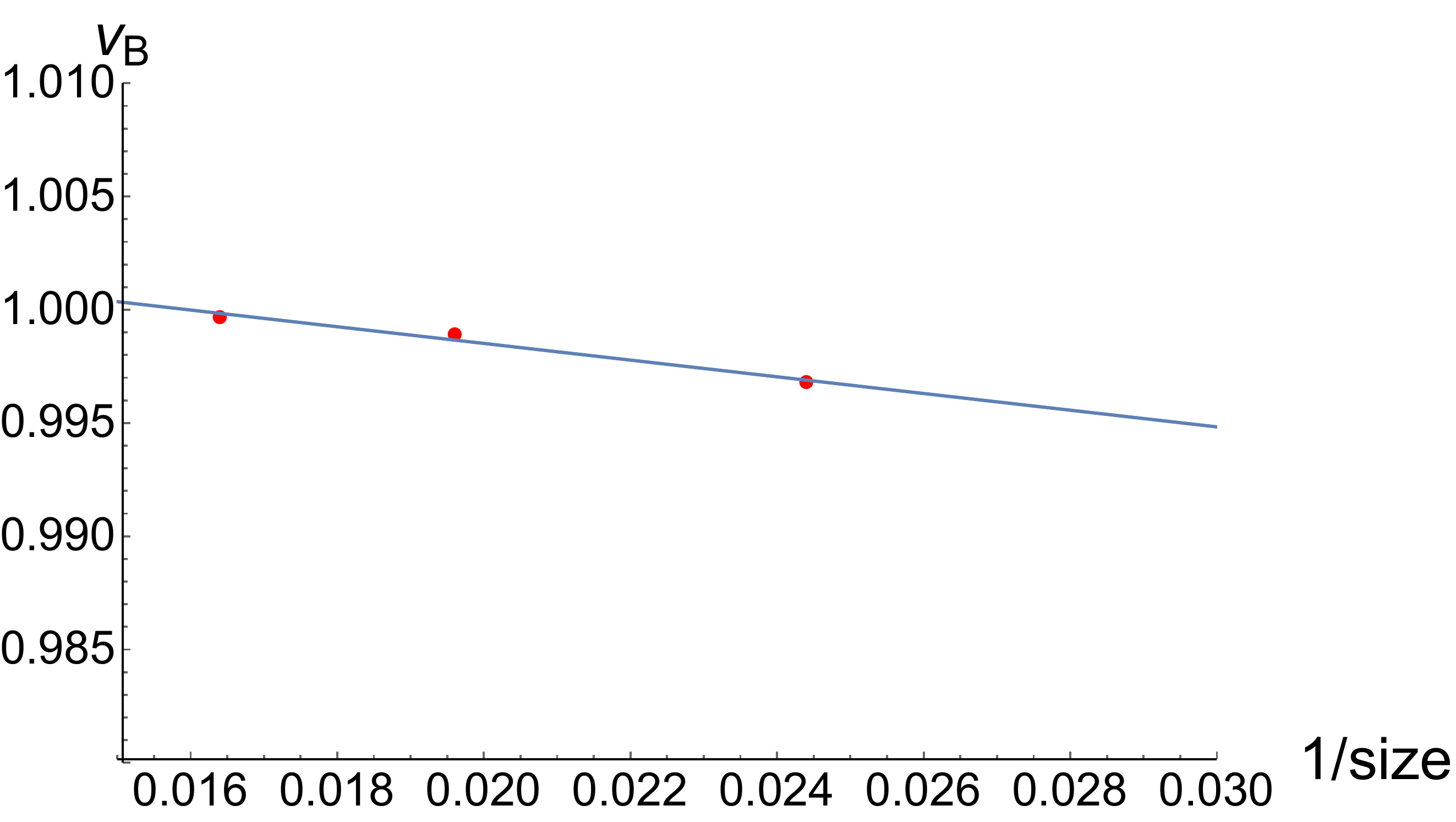}
  \caption{Butterfly velocity versus grid size computed without assuming rotational invariance. Also shown is a linear fit to the data. The extrapolated value is slightly bigger than one, which violates causality, so we simply say that $v_B/c \approx 1$. The largest value obtained in the calculation (which occurs for the largest grid size considered) is $v_B/c \approx .999$.}
  \label{fig:Vk}
\end{figure}

Our numerical computation of $v_B$ proceeds as in Section \ref{LE} except that we no longer assume rotational invariance. This is because the leading eigenvector is not rotationally invariant at finite external momentum $\k$. We approximately compute the largest eigenvalue of the integral kernel $\hat{\mathcal{K}}$ as a function of momentum $\k$ and find excellent agreement with the form postulated in Eq.~\eqref{lambdaquad}. From these data we extract $\lambda_L$ and $v_B$; we also extract $1/\tau_\varphi$. The data are shown in Figures \ref{fig:Gk}, \ref{fig:Lk}, and \ref{fig:Vk}.

Curiously, the extrapolated value of $v_B/c$ is very slightly greater than one, but of course we must have $v_B/c \leq 1$ by causality. However, the largest value obtained in the calculation for the largest grid size considered by us is $v_B /c \approx .999$. In fact, if the form of $\mathcal{C}$ in Eq.~\eqref{finctx} is valid, we claim that $v_B/c$ should be strictly less than one. This is because Eq.~\eqref{finctx} does not exactly vanish outside the ``butterfly cone" $|\x| = v_B t$. Since the commutator must exactly vanish outside the light cone $|\x| = ct$, it follows that either Eq.~\eqref{finctx} is incorrect or $v_B < c$. All our numerical results are consistent with the basic form Eq.~\eqref{finctx}, so we argue that $v_B < c$. 

\section{Onset of chaos in proximate phases}
\label{proximate}

Although we leave the full quantitative calculations to future work, the framework developed above enables a qualitative discussion of the behavior of $\lambda_L$ and $v_B$ in the symmetry broken and unbroken phases separated by the quantum critical point. We focus only on the low temperature limit inside each phase; more generally there is a complicated crossover function which describes the entire phase diagram but is beyond the scope of this work.

First consider the symmetry unbroken phase. Here the $\varphi$ particles have a finite mass, so at the lowest temperatures the system will correspond to a dilute gas of weakly interacting well-defined $\varphi$ quasiparticles. The scattering rate (or inverse dephasing time) will be proportional to the square of the particle density since two particles must approach close to each other in order to scatter. Assuming the chaos exponent is comparable to the scattering rate (as we found above), we presumably have $\lambda_L \sim e^{- 2 \mu/T}$ where $\mu\sim\sqrt{|g-g_c|}$ is the mass of $\varphi$. Such a dilute gas of $\varphi$ particles has an interparticle spacing much larger than the thermal wavelength, hence the gas is effectively classical. Expanding the dispersion $\epsilon_{\k}$ near zero momentum gives $\epsilon_{\k} = \mu + |\k|^2/2\mu$, so equipartition of energy gives a typical speed of the particles of order $\sqrt{T/\mu}$. Keeping in mind the order of limits issue discussed in Section \ref{VB}, we expect that at the longest times there is a butterfly velocity going like $v_B \sim \sqrt{T/\mu}$. This is also consistent with the scaling form of $v_B$ with $z=2$.

Now consider the symmetry broken phase. Our analysis here is schematic since the symmetry broken phase does not survive to non-zero temperature in two spatial dimensions. The symmetry broken phase can be stabilized at non-zero temperature by considering a three dimensional system, e.g. weakly coupled layers of the two dimensional system.

The dominant low energy degrees of freedom are the massless Goldstone modes denoted $\pi_a$, $a=1,...,N-2$ (the amplitude mode is gapped and by the discussion in the previous paragraph, leads to slower growth of chaos at low temperatures). However, in addition to being massless, they are also derivatively coupled which will suppress the growth of chaos at the lowest temperatures. For $N>2$ (the non-linear sigma model) the leading correction to the free Goldstone action is determined just by symmetry,
\beq
\mathcal{L} \sim \rho_s \left[ (\partial \pi_a)^2 + (\pi_a \partial \pi_a)^2 - \pi^2_a (\partial \pi_b)^2 + ...\right],
\eeq
where $\rho_s$ is the stiffness. The dimension of $\rho_s$ is $[\rho_s]_2 = 1$ in two spatial dimensions.

Introducing the canonically normalized field $\chi_a = \sqrt{\rho_s} \pi_a$, we see that the second term is an irrelevant interaction with a universal coefficient determined by $\rho_s$. Assuming the basic rung contains two factors of the interaction vertex, dimensional analysis then suggests that in two dimensions the chaos exponent goes like $\lambda_{L,2} \sim T^3 / \rho_s^2$. The Goldstone bosons are massless and hence move at the speed of light, however, by analogy with the Boltzmann-like equation derived above, the butterfly velocity may in general be smaller than but of the same order as the speed of light.

The case $N=2$ is special since the leading symmetry controlled irrelevant operator vanishes. The leading correction to free Goldstone action, which is $\rho_s (\partial \theta)^2$, is now determined by non-universal physics via the operator $g (\partial \theta)^4$. In two spatial dimensions $g$ has energy dimension $[g]=-1$. As before, introducing $\chi = \sqrt{\rho_s} \theta$ gives a canonically normalized free scalar with an irrelevant interaction $(g/\rho_s^2) (\partial \chi)^4$. Dimensional analysis now suggests that the chaos exponent is of order $g^2 T^7/\rho_s^4$. However, we still expect a finite butterfly velocity of the order of the speed of light.

However, this analysis of the $N=2$ case is complicated by the presence of vortex excitations in two dimensions. Unlike for $N>2$, the physics of the symmetry broken phase is not completely destroyed at non-zero temperature. Phase fluctuations of the complex $N=2$ order parameter do destroy the long-range order, but there is a distinct finite temperature phase in which vortices have not proliferated.

For completeness, we record our expectations for the three dimensional case where the symmetry broken state is stable at small non-zero temperature. The dimension of $\rho_s$ is now $[\rho_s]_3 = 2$ in three spatial dimensions. Again introducing a canonically normalized scalar field and assuming that the basic rung has two copies of the interaction vertex, dimensional analysis gives the estimate $\lambda_{L,3} \sim T^5/\rho_s^2$ and $v_B \sim c$.

\section{Discussion}

In this work we studied the onset of chaos in the $O(N)$ nonlinear sigma model in a controlled $1/N$ expansion. As the model is weakly interacting at infinite $N$, we found a chaos exponent $\lambda_L = 3.2~T/N$ which vanished in the limit of $N\rightarrow\infty$ and trivially obeys the chaos bound. The calculation predicts that models of smaller $N$ scramble faster and, making an uncontrolled extrapolation all the way to $N=1$, suggests that the Ising quantum critical point scrambles with a growth exponent $\lambda_L^{(N=1)} \approx 3.2~ T$. We also studied the butterfly velocity and found that at weak coupling the butterfly velocity (in units of the speed of light) is a universal number independent of $T$ and $N$ at large $N$. At finite $N$ we expect $v_B$ to remain independent of $T$ but to potentially depend on $N$. The results for the $O(N)$ model can be contrasted with a different large $N$ problem of a Fermi surface of $N$ component fermions coupled to a $U(1)$ gauge field, where $\lambda_L$ is also proportional to $T$ but not suppressed by $1/N$ (although the precise numerical prefactor is difficult to reliably compute due to the breakdown of the $1/N$ expansion which removes the $1/N$ suppression in $\lambda_L$) and $v_B \sim T^{1/3}$ \cite{PatelSachdev}.

Our calculation is uncontrolled at small $N$, which includes the most physically relevant cases, so in this case it would be quite interesting to carry out experimental measurements in suitable model systems. It would also be interesting to carry out a complementary $\epsilon=3-d$ expansion at a fixed $N$ in order to study scrambling at the Wilson-Fisher conformal fixed point \cite{QPT}.

Recent work has also focused on the possible relation between diffusion, the chaos exponent, and the butterfly velocity \cite{BSDC17} \cite{Blake16} \cite{GuQiStanford}. In particular, a generalized version \cite{GuQiStanford} of the Sachdev-Ye-Kitaev (SYK) \cite{SY} \cite{kitaevtalk} finds an exact relationship between the energy diffusion constant, $D_E = v_B^2 / \lambda_L$, while no such relationship exists for the charge diffusion constant, $D_c$ \cite{SS16}.  The energy diffusion constant is undefined at the $O(N)$ quantum critical point due to an absence of momentum relaxation. If the stress tensor is given by $T^{\mu \nu}$, then the energy current $T^{0i}$ is also the momentum density. Hence the total energy current cannot relax unless momentum is also not conserved. However, if we choose a ``charge" $U(1)$ subgroup of the $O(N)$ symmetry, then the charge (for $N=2$), or, spin (for $N=3$) diffusion constant is well defined and known at large $N$.

The Einstein relation is $D_c = \sigma_{\textnormal{dc}}/\chi$ where the charge susceptibility is $\chi = \sqrt{5} \Theta T/2\pi$ \cite{QPT} and the dc conductivity is $\sigma_{\textnormal{dc}} = 0.085 N$ \cite{QPT} \cite{WWKTS12}. The Einstein relation then gives $D_c = 0.249 N/T$. By comparison, we find that $v_B^2 / \lambda_L \approx 0.31 N/T$ which is reasonably close to the value for $D_c$. Note, however, that in two spatial dimensions so-called long-time tails ultimately lead to divergences in transport coefficients \cite{Pomeau75} \cite{WWKSS14}, although this effect is suppressed at large $N$. Hence in the scaling limit the true diffusion constant is infinite while $v_B^2/\lambda_L$ remains finite. It continues to be an interesting problem to explore the extent to which charge and energy diffusion constants in generic strongly correlated systems without a well-defined quasiparticle description are related to quantum chaos. If any relationship exists, this would be interesting for studying transport in various `strange' metallic phases that are ubiquitous in correlated materials, such as the cuprates, the pnictides and other heavy-fermion materials \cite{Mackenzie13}.

Given that the calculation focused on the elementary field operators, one may also wonder about the extent to which commutators of other operators grow at the same rate. That commutators of other operators cannot grow more slowly is easily seen. To illustrate the physics, consider the example of the composite operator $\varphi^2$; in the nonlinear sigma model the field $\lambda$ plays the role $\varphi^2$. Consider then
\beq
\mathcal{C}_{\varphi^2} = \text{Tr}\left(\sqrt{\rho} [\varphi^2(\x,t),\varphi^2(\vec{0},0)]\sqrt{\rho} [\varphi^2(\x,t),\varphi^2(\vec{0},0)]  \right).
\eeq
The commutator satisfies
\beq
[\varphi^2(\x,t),\varphi^2(\vec{0},0)] =  \varphi(\x,t)[\varphi(\x,t),\varphi(\vec{0},0)] \varphi(\vec{0},0) + ...
\eeq
where $...$ denotes three other terms with different placements of the operators. In perturbation theory we can then contract the extra factors of $\varphi$ to leave the original expression for $\mathcal{C}$, i.e.
\begin{align}
\mathcal{C}_{\varphi^2} &= \langle \varphi(\x,t) \varphi(\vec{0},0)\rangle \langle \varphi(\x,t) \varphi(\vec{0},0)\rangle\, \mathcal{C} \nonumber \\
&+ \langle \varphi(\x,t) \varphi(\x,t)\rangle \langle \varphi(\vec{0},0) \varphi(\vec{0},0)\rangle\, \mathcal{C} + ...
\end{align}
where $...$ denotes contributions from additional diagrams that don't factorize. Note also that the first term will be exponentially small at finite temperature if $|\x| \gg \beta$, so only the second term contributes, but this term amounts to $\mathcal{C}$ up to a local renormalization equal to $\langle \varphi^2 \rangle^2$. Assuming the other $...$ contributions do not magically cancel this term, we have shown that $\mathcal{C}_{\varphi^2}$ grows as fast as $\mathcal{C}$ (note, however, that the time for $\mathcal{C}_{\varphi^2}$ to become large may differ owing to the renormalization).

In diagrammatic terms, given any set of composite operators, we can always dress a copy of the elementary $\varphi$ ladder with some local renormalizations (contractions between operators at the same spacetime point) to obtain a set of diagrams that yield an exponentially growing composite operator square commutator. Thinking more broadly about weakly coupled theories, this general argument suggests that commutators of composite operators should grow as fast as commuators of elementary operators (fields that give a good description of the unperturbed theory). What this general argument does not show is that commutators of composite operators must grow at the same rate as commutators of elementary operators. One could worry that they could potentially grow faster, e.g. due to some special collective modes that were not included in the ``elementary ladder". However, this is unlikely on physical grounds. Diagrammatically, we can imagine inserting the ``composite ladder" into the diagrammatic expansion for the commutator of elementary fields, thus showing that they actually grow at the same rate. Physically, products of local composite operators generate sums over local operators, so generic sets of operators are expected to all have commutators which grow at the same rate. In short, chaos in one local operator easily spreads to all other local operators. Possible exceptions to this rule include non-local operators like defect operators or line operators.

Finally, the computations presented here can be extended to a variety of other models, including to fermionic QED$_3$ in (2+1)-dimensions \cite{qed3} and to the weak coupling Banks-Zaks fixed point in (3+1) dimensions \cite{bankszak}. It will be interesting to understand, for example, if a more highly entangled critical point like the one described by QED$_3$ scrambles faster than the corresponding $O(N)$ critical point \cite{toappear}.

\section{Acknowledgements}
We thank J. McGreevy, A.A. Patel, S. Sachdev, D. Stanford, J. Steinberg, W. Witczak-Krempa and L. Zou for valuable disussions. We especially thank W. Witczak-Krempa and S. Sachdev for helping benchmark some of our numerical computations of the phase-coherence time and L. Zou for a careful reading of our manuscript. DC is supported by a postdoctoral fellowship from the Gordon and Betty Moore Foundation, under the EPiQS initiative, Grant GBMF-4303, at MIT. BGS is supported by the Simons Foundation as part of the It From Qubit collaboration and through a Simons Investigator Award to Senthil Todadri and by a MURI grant W911NF-14-1-0003 from ARO.

\appendix
\section{Thermal mass}
\label{mass}
Here we review the standard computation for the thermal contribution to the mass, $\mu$, at the critical point $g=g_c$. In Eq.~\eqref{HSl}, and at infinite $v$, the $\lambda$ field exactly implements the hard constraint
\beq
\varphi_a^2 = \frac{N}{g}.
\eeq
At large $N$, the constraint is implemented on average and so we may replace this hard constraint by its expectation value,
\beq
N T\sum_{i\omega_n} \int_\k \frac{1}{\omega_n^2 + \k^2 + \mu^2} = \frac{N}{g}.
\eeq
It is convenient to adopt a regulator scheme where we keep all frequencies but implement a hard momentum cutoff $\Lambda$. The quantum critical point is then determined by setting $\mu^2 =0$:
\beq
\frac{1}{g_c} = \int^\Lambda_\k \frac{1}{2|\k|}.
\eeq
For $d=2$ the above implies,
\beq
\frac{1}{g_c} = \frac{\Lambda}{4 \pi}.
\eeq
The thermal mass is then determined by
\beq
\frac{1}{g_c} = \int^\Lambda_\k \frac{1}{2 \epsilon_\k }\tanh \frac{\beta \epsilon_\k}{2}.
\eeq
The analysis is easiest if we add and subtract the zero temperature integral but with $\mu^2$ restored:
\beq
\int^\Lambda_{\k} \frac{1}{2|\k|} - \int^\Lambda_{\k} \frac{1}{2\epsilon_{\k}} = \int^\Lambda_{\k} \frac{1}{e^{\beta \epsilon_{\k}} -1}\frac{1}{\epsilon_{\k}}.
\eeq
Both the LHS and the RHS are manifestly UV convergent; the result is a non-linear equation for $\mu = \Theta T$.

The equation for the thermal mass $\mu = \Theta T$ is (the integral is done by changing variables to $\epsilon/T$)
\beq
\frac{\Theta T}{4\pi} = \frac{T}{2\pi} \ln\(\frac{e^\Theta}{e^\Theta -1} \).
\eeq
$\Theta$ thus obeys
\beq
e^{\Theta/2} = \frac{e^\Theta}{e^\Theta -1},
\eeq
with the usual physical solution
\beq
\Theta = 2 \ln\( \frac{1+\sqrt{5}}{2}\) \approx 0.962.
\eeq

\section{One-loop bubble}
\label{oneloop}
Here we evaluate the $\varphi_a$ bubble:
\beq
\Pi(i \nu_n,\q) = \frac{T}{2} ~\sum_{\omega_n} \int^\Lambda_\k \frac{1}{(\omega_n + \nu_n)^2 + \epsilon_{\k+\q}^2}\frac{1}{\omega_n^2 + \epsilon^2_\k},
\eeq
where $\int^\Lambda_\k\equiv\int_{|\k|<\Lambda}\frac{d^d \k}{(2\pi)^d}$.
We first rewrite the bubble as,
\beq
\Pi(i \nu_n,\q) = \frac{1}{2} \oint\frac{dz}{2\pi i} \int^\Lambda_\k ~b(z) \frac{1}{(z + i\nu_n)^2 - \epsilon^2_{\k+\q}}\frac{1}{z^2 - \epsilon^2_\k},
\eeq
where $b(z)=(e^{\beta z}-1)^{-1}$. Evaluating the residues and simplifying the final expressions, we obtain,
\beq
\Pi(i \nu_n,\q) = - \frac{1}{2} \int^\Lambda_\k \frac{1}{4\epsilon_\k \epsilon_{\k+\q}} \bigg[\frac{b(\epsilon_{\k+\q})-b(\epsilon_\k)}{\epsilon_{\k+\q}-\epsilon_\k + i\nu_n} &+& \frac{b(\epsilon_\k)-b(\epsilon_{\k+\q})}{\epsilon_\k-\epsilon_{\k+\q} + i\nu_n} \nonumber\\
-\frac{b(\epsilon_{\k+\q})-b(-\epsilon_\k)}{\epsilon_{\k+\q} + \epsilon_\k - i\nu_n} &-& \frac{b(\epsilon_\k)-b(-\epsilon_{\k+\q})}{\epsilon_\k + \epsilon_{\k+\q} + i\nu_n}\bigg].
\eeq
The overall minus sign arises because the non-Matsubara poles are encircled clockwise instead of counterclockwise.

The imaginary part of the retarded bubble after analytically continuing to real frequencies $(i\nu_n\rightarrow \nu+i0^+)$ is,
\beq \label{imbformula}
\textnormal{Im}[\Pi_R(\nu+i0^+,\q)] =  \frac{1}{2} \int^\Lambda_\k \frac{\pi}{4\epsilon_\k \epsilon_{\k+\q}} \bigg[[b(\epsilon_{\k+\q})-b(\epsilon_\k)]\delta(\epsilon_{\k+\q}-\epsilon_\k+\nu) \nonumber\\+ [b(\epsilon_\k)-b(\epsilon_{\k+\q})]\delta(\epsilon_\k-\epsilon_{\k+\q}+\nu) \nonumber\\
+[b(\epsilon_{\k+\q})-b(-\epsilon_\k)]\delta(\epsilon_{\k+\q}+\epsilon_\k-\nu) - [b(\epsilon_\k)-b(-\epsilon_{\k+\q})]\delta(\epsilon_\k+\epsilon_{\k+\q}+\nu)\bigg].
\eeq
We note that the above is an odd function of $\nu$, as should be the case. It is positive for positive frequency $\nu$.

As a simple check, let us evaluate the bubble in the zero momentum limit, $\q\rightarrow0$. This limit yields
\beq
\textnormal{Im}[\Pi_R(\nu+i0^+,\q=0)] =  \frac{1}{2} \int^\Lambda_\k \frac{\pi}{4\epsilon_\k^2} \bigg[
[1+2b(\epsilon_\k)]\delta(2\epsilon_\k-\nu) - [1+2b(\epsilon_\k)]\delta(2\epsilon_\k+\nu)\bigg].
\eeq
In two spatial dimensions, $d=2$, we find
\beq
\textnormal{Im}[\Pi_R(\nu+i0^+,\q=0)] &=&  \int d\epsilon~ \frac{1}{16\epsilon} \bigg[
[1+2b(\epsilon)]\delta(2\epsilon-\nu) - [1+2b(\epsilon)]\delta(2\epsilon+\nu)\bigg],
\eeq
\beq
\textnormal{Im}[\Pi_R(\nu+i0^+,\q=0)] &=& \frac{1}{16\nu}[1+2b(\nu/2)]\theta\bigg(\frac{\nu}{2}-\mu\bigg) \nonumber\\
&& + \frac{1}{16\nu}[1+2b(-\nu/2)]\theta\bigg(-\frac{\nu}{2}-\mu\bigg).
\eeq
This matches nicely with what we would expect from the Cutkosky cutting rules. This result can be compared with the zero temperature result,
\beq
\Pi_{T=0}(i\nu_n,\q) = \frac{1}{16} \frac{1}{\sqrt{|\q|^2 + \nu_n^2}}.
\eeq
Continuing to real frequencies, $i\nu_n \rightarrow \nu + i 0^+$, the imaginary part of $\Pi_{T=0}$ is non-zero for $\nu > q$. We have the standard result that
\beq
\text{Im}[\Pi_{T=0}] = \frac{1}{16} \frac{1}{\sqrt{\nu^2 - |\q|^2}} \theta(\nu - |\q|).
\eeq
In the limit that $\nu \gg T$, the zero temperature and finite temperature results agree, e.g. as can be seen from the finite temperature formula at $\q=0$.

Returning to the general case, since $\textnormal{Im}[\Pi_R(\nu+i0^+,q)]$ is odd in $\nu$, we only have to evaluate it for $\nu>0$. In that case, we can drop the last term in the expression Eq.~\eqref{imbformula} for $\textnormal{Im}[\Pi_R(\nu+i0^+,q)]$ since the argument in the delta function is positive. Moreover, rotational invariance implies that the first term is equal to the second term, so we find
\beq
\textnormal{Im}[\Pi_R(\nu+i0^+,\q)]_{\nu>0} = \frac{1}{2} \int^\Lambda_\k \frac{\pi}{4\epsilon_\k\epsilon_{\k+\q}} \bigg[2[b(\epsilon_{\k+\q})-b(\epsilon_\k)]\delta( \nu + \epsilon_{\k+\q}-\epsilon_\k) \nonumber\\
+[b(\epsilon_{\k+\q})-b(-\epsilon_\k))]\delta(\epsilon_{\k+\q}+\epsilon_\k-\nu)\bigg].
\eeq

In order to evaluate the above quantity numerically, it is useful to recast it as follows. Shift the $\k$ integral so that the energies are
\beq
\epsilon_\pm = \epsilon_{\k \pm \q/2}.
\eeq
The first delta function requires
\beq
\epsilon_+ - \epsilon_- + \nu =0
\eeq
while the second requires
\beq
\epsilon_+ + \epsilon_- - \nu = 0.
\eeq
These may be put in the common form
\beq
\epsilon_+ + s \nu = s \epsilon_-
\eeq
where $s = \pm 1$. Squaring both sides gives
\beq
\epsilon_+^2 + 2 s \nu \epsilon_+ + \nu^2 = \epsilon_-^2.
\eeq

In radial coordinates $\{|\k|,\theta\}$, the energies are
\beq
\epsilon_\pm = \epsilon_{\k \pm \q/2} = \sqrt{|\k|^2 \pm |\k| |\q| \cos \theta + \q^2/4 + \mu^2}.
\eeq
The above condition then translates to
\beq
|\k| = |\k|_0 =  \frac{\nu}{2} \sqrt{\frac{\nu^2 - |\q|^2 - 4 \mu^2}{\nu^2 - |\q|^2 \cos^2 \theta}}.
\eeq
Note that the integral is only over positive $|\k|$, so there is only one root to consider. Furthermore, in order to have $|\k|_0$ real, we must have $\nu^2 < |\q|^2 \cos^2 \theta$ or $\nu^2 > |\q|^2 + 4 \mu^2$. Finally, to actually satisfy the delta function, we must check that $\epsilon_+ + s\nu = s \epsilon_-$. For $\nu < q$ the only solution is for $s=1$ in the range $\theta \in (\pi - \cos^{-1}\frac{\nu}{q},\pi + \cos^{-1}\frac{\nu}{q})$. For $\nu > q^2+ 4 \mu^2$ the only solution is for $s=-1$ in the range $\theta \in (0, 2\pi)$.

\section{Wightman function}
\label{wight}

In order to express the symmetric Wightman function, $\CG_W(\omega,\k)$ in terms of the spectral function, we first insert a complete set of (many-body) eigenstates, $|n\rangle$ (with energy $E_n$ and momentum $\P_n$)
\beq
\CG_W(\omega,\k) = \int dt ~e^{i \omega t} \sum_{n,m} \frac{e^{-\beta (E_n+E_m)/2}}{Z} |\varphi_{mn}|^2 ~\delta^d(\k - (\P_m - \P_n)) e^{-i(E_m - E_n)t},
\eeq
where $Z$ is the partition-function and $|\varphi_{mn}|^2=|\langle m|\varphi|n\rangle^2$.

Now recall that for $\CG_R(\omega,\k)$ a similar procedure leads to
\beq
\CG_R(\omega,\k) = \sum_{n,m} \frac{e^{-\beta E_n}}{Z} |\varphi_{mn}|^2 \left\{ \frac{\delta^d(\k - (\P_m - \P_n))}{\omega + i 0^+ - (E_m - E_n)} -   \frac{\delta^d(\k + (\P_m - \P_n))}{\omega + i0^+ + (E_m - E_n)}\right\},
\eeq
and so we can read off the spectral function as,
\beq
A(\omega,\k) = \sum_{n,m} 2\pi \frac{e^{-\beta E_n}}{Z} |\varphi_{mn}|^2 \{\delta^d(\k - (\P_m - \P_n)) ~\delta(\omega - (E_m-E_n)) \nonumber\\
 -  \delta^d(\k + (\P_m - \P_n)) \delta(\omega + (E_m-E_n))\}.
\eeq

The above can be simplified further by using the delta-function constraint,
\beq
A(\omega,\k) &=& \sum_{n,m} 2\pi \frac{e^{-\beta E_n}}{Z} |\varphi_{mn}|^2 \delta^d(\k - (\P_m - \P_n))~ \delta(\omega - (E_m-E_n)) (1 - e^{-\beta \omega}),\\
A(\omega,\k) &=& \sum_{n,m} 2\pi \frac{e^{-\beta (E_n+E_m)/2}~e^{\beta\omega/2}}{Z} |\varphi_{mn}|^2 \delta^d(\k - (\P_m - \P_n))~ \delta(\omega - (E_m-E_n)) (1 - e^{-\beta \omega}).\nonumber\\
\eeq

Therefore, the Wightman function can be expressed as,
\beq
\CG_W(\omega,\k) = ~ \frac{e^{-\beta \omega/2}}{1-e^{-\beta\omega}} A(\omega,\k)= \frac{A(\omega,\k)}{e^{\beta \omega/2} - e^{-\beta \omega/2}},
\eeq
which leads to the useful relation
\beq
\CG_{W}(\omega,\k) = \frac{A(\omega,\k)}{2\sinh(\beta \omega/2)}.
\label{gwa}
\eeq

\section{One-loop self-energy}
\label{SE}
Here we evaluate the self-energy of $\varphi_a$:
\beq
\tilde\Sigma(i \omega_n, \q) = \frac{T}{N} \sum_{i\nu_n} \int^\Lambda_\k ~\bigg[\CG(i \omega_n + i \nu_n, \q+\k) - \CG(i \nu_n,\k)\bigg]~\CG_\lambda(i \nu_n,\k).
\label{imsigma_app}
\eeq

Let us focus on the contribution from the first term above, which we denote $\tilde\Sigma_1(i \omega_n, \q)$,
\beq
\tilde\Sigma_1(i\omega_n,\q) = \frac{T}{N}  \sum_{i\nu_n} \int^\Lambda_\k \frac{1}{(\omega_n+\nu_n)^2+\epsilon^2_{\k+\q}}~\CG_\lambda(i \nu_n,\k).
\eeq

We can re-express the above as
\beq
\tilde\Sigma_1(i\omega_n,\q) = -\frac{1}{N}\oint\frac{dz}{2\pi i} \int^\Lambda_\k ~b(z)~\frac{1}{(z+i\omega_n)^2-\epsilon^2_{\k+\q}}~\CG_\lambda(z,\k).
\eeq
Evaluating this leads to,
\beq
\tilde\Sigma_1(i\omega_n,\q) &=& -\frac{1}{N}\int^\Lambda_\k \bigg[\bigg(\int_{-\infty}^\infty \frac{dx}{2\pi i}~\frac{b(x)}{(x+i\omega_n)^2-\epsilon^2_{\k+\q}}~[\CG_\lambda(x+i0^+,\k)-\CG_\lambda(x-i0^+,\k)]\bigg)\nonumber\\
&-& \frac{b(\epsilon_{\k+\q})}{2\epsilon_{\k+\q}}\CG_\lambda(\epsilon_{\k+\q}-i \omega_n,\k) -  \frac{1+b(\epsilon_{\k+\q})}{2\epsilon_{\k+\q}}\CG_\lambda(-\epsilon_{\k+\q}-i \omega_n,\k) \bigg].
\eeq
The minus signs in front of the second two terms again arise from the clockwise encircling of the poles.

Let us shift $\k\rightarrow \k-\q$ and further simplify the above expressions,
\beq
\tilde\Sigma_1(i\omega_n,\q) = -\frac{1}{N}\int^\Lambda_\k \bigg[\bigg(\int_{-\infty}^\infty \frac{dx}{\pi}~\frac{b(x)}{(x+i\omega_n)^2-\epsilon^2_{\k}}~\tn{Im}[-\CG_{R,\lambda}(x,\k-\q)]\bigg)\nonumber\\
- \frac{b(\epsilon_\k)}{2\epsilon_\k}\CG_\lambda(\epsilon_\k-i \omega_n,\k-\q) -  \frac{1+b(\epsilon_\k)}{2\epsilon_\k}\CG_\lambda(-\epsilon_\k-i \omega_n,\k-\q) \bigg].
\eeq
We have used the formula Eq.~\eqref{lambdaeuctoret} relating the Euclidean $\lambda$ correlator with the retarded correlator (note the minus sign). The analytic continuation is $i\omega_n\rightarrow \omega+i0^+$ and the imaginary part is
\beq
\tn{Im}[\tilde\Sigma_1(\omega+i0^+,\q)] = \frac{1}{N}\int^\Lambda_\k \bigg[\bigg(\int_{-\infty}^\infty dx~\frac{b(x)}{2\epsilon_\k}~[-\delta(x+\omega-\epsilon_\k)+\delta(x+\omega+\epsilon_\k)]~\tn{Im}[\CG_{R,\lambda}(x,\k-\q)]\bigg)\nonumber\\
+ \frac{b(\epsilon_\k)}{2\epsilon_\k}\tn{Im}[\CG_{R,\lambda}(\epsilon_\k- \omega,\k-\q)] +  \frac{1+b(\epsilon_\k)}{2\epsilon_\k}\tn{Im}[\CG_{R,\lambda}(-\epsilon_\k- \omega,\k-\q)] \bigg],\nonumber\\
\eeq
The plus signs in the last two terms arise because $\text{Im}[\CG_\lambda(x-i0^+)] = - \text{Im}[\CG_\lambda(x+i0^+)]$. The overall positive sign comes from the relative sign relating $\CG_\lambda$ and $\CG_{R,\lambda}$.

Further simplification yields
\beq
\tn{Im}[\tilde\Sigma_1(\omega+i0^+,\q)] = \frac{1}{N}\int^\Lambda_\k \bigg[-\frac{b(\epsilon_\k-\omega)}{2\epsilon_\k}~\tn{Im}[\CG_{R,\lambda}(\epsilon_\k-\omega,\k-\q)] \nonumber\\
+ \frac{b(-\epsilon_\k-\omega)}{2\epsilon_\k}~\tn{Im}[\CG_{R,\lambda}(-\epsilon_\k-\omega,\k-\q)]\nonumber\\
+ \frac{b(\epsilon_\k)}{2\epsilon_\k}\tn{Im}[\CG_{R,\lambda}(\epsilon_\k- \omega,\k-\q)] +  \frac{1+b(\epsilon_\k)}{2\epsilon_\k}\tn{Im}[\CG_{R,\lambda}(-\epsilon_\k - \omega,\k-\q)] \bigg].
\eeq
Finally,
\beq
\tn{Im}[\tilde\Sigma_1(\omega+i0^+,\q)] = \frac{1}{N}\int^\Lambda_\k \bigg[
&&\frac{b(\epsilon_\k)-b(\epsilon_\k-\omega)}{2\epsilon_\k}\tn{Im}[\CG_{R,\lambda}(\epsilon_\k- \omega,\k-\q)] \nonumber\\
+&&  \frac{b(\epsilon_\k)-b(\epsilon_\k+\omega)}{2\epsilon_\k}\tn{Im}[\CG_{R,\lambda}(-\epsilon_\k- \omega,\k-\q)] \bigg].
\label{final1}
\eeq

It should be clear from the structure of the above expression that $\tn{Im}[\tilde\Sigma] = \tn{Im}[\tilde\Sigma_1]$, i.e. the second piece in Eq.~\eqref{imsigma_app} doesn't contribute to the imaginary part.

It is useful to go one step further and express the self-energy in terms of the Wightman function. Eq.~ \eqref{final1} then becomes,
\beq
\tn{Im}[\tilde\Sigma(\omega+i0^+,\q)] = -\frac{1}{N}\int^\Lambda_\k \frac{\sinh(\beta\omega/2)}{4\epsilon_\k\sinh(\beta\epsilon_\k/2)}\bigg[
&&\frac{1}{\sinh[\beta(\epsilon_\k-\omega)/2]}\tn{Im}[\CG_{R,\lambda}(\epsilon_\k- \omega,\k-\q)] \nonumber\\
+&& \frac{1}{\sinh[-\beta(\epsilon_\k+\omega)/2]}\tn{Im}[\CG_{R,\lambda}(-\epsilon_\k - \omega,\k-\q)] \bigg].\nonumber\\
\eeq

Recall that the spectral function,
\beq
A(\omega,\k) = -2~\tn{Im}[\CG_R(\omega,\k)],
\eeq
and using the relationship between the Wightman function and the spectral function in Eq.~\eqref{gwa}, we have
\beq
\tn{Im}[\Sigma(\omega+i0^+,\q)] =   \frac{1}{N}\int^\Lambda_\k \frac{\sinh(\beta\omega/2)}{4\epsilon_\k\sinh(\beta\epsilon_\k/2)}\bigg[\CG_{W,\lambda}(\epsilon_\k-\omega,\k-\q) + \CG_{W,\lambda}(-\epsilon_\k-\omega,\k-\q)\bigg].\nonumber\\
\eeq
Tracking through the minus signs, we see that the imaginary part of the self-energy is positive for positive frequency.

\section{Feynman rules for ladder diagrams}
\label{sec:diagramrules}

Here we show how the diagrammatic rules for performing the ladder sum are derived. We focus on the ``bare" ladder; the dressing of propagators due to self-energy effects is standard and will not be reviewed here. The basic object of interest is again the squared commutator ($\varphi(0) \equiv \varphi(\vec{0},0)$):
\beq \label{regcommapp}
\mathcal{C}(\x,t) = -\frac{1}{N^2} \sum_{a,b} \tr\left\{ \rho^{1/2} [\varphi_a(\x,t),\varphi_b(0)] \rho^{1/2} [\varphi_a(\x,t),\varphi_b(0)] \right\}.
\eeq

Recall that at infinite $N$ the fields $\varphi_a$ are free, so $\mathcal{C}$ reduces to a free particle result. There are two sources of $1/N$ corrections to the $N\rightarrow\infty$ form of $\mathcal{C}$, one from the thermal state $\rho$ and one from the time-evolution operator $e^{- i H t}$ defining the Heisenberg operators in Eq.~\eqref{regcommapp}. As discussed in the main text, we may ignore the corrections to the thermal state when computing the chaos exponent to leading order in $1/N$. The nontrivial contributions to scrambling thus arise from the time-evolution operator.

It is convenient to go to the {\it interaction} picture with respect to the non-interacting $N \rightarrow \infty$ Hamiltonian. Then the non-trivial part of the time evolution operator is given by a time-ordered exponential of the interaction term in the interaction picture:
\beq
- \frac{1}{2\sqrt{N}} \sum_a \int_0^t ds \int_{\x} \lambda_{0}(\x,s) \varphi^2_{a,0}(\x,s)
\eeq
(the overall minus sign in the Hamiltonian arises because the interaction has a positive sign in the Lagrangian). The subscript `$0$' indicates that these fields evolve under the non-interacting Hamiltonian. To be precise, the interaction picture time-evolution operator is
\beq
U_I = \mathcal{T} \exp\left( \frac{i}{2\sqrt{N}}\sum_a \int_0^t ds \int_{\x} \lambda_0(\x,s) \varphi^2_{a,0}(\x,s)\right).
\eeq

For clarity of presentation, we henceforth drop the factors of $N$ and the index structure. These can be easily restored using the rules discussed in the main text. Our purpose in this appendix is to understand the overall sign of diagrams, the various combinatorial factors, and the analytic structure i.e. which Greens functions appear.

We may thus rewrite $\mathcal{C}$ as (after dropping factors of $N$ and various indices)
\beq
\mathcal{C}(\x,t) = -\tr\left\{ \rho^{1/2} [U_I^\dagger \varphi_0(\x,t) U_I ,\varphi_0(0)] \rho^{1/2} [U_I^\dagger \varphi_0(\x,t) U_I ,\varphi_0(0)] \right\}.
\eeq
This formula is the basis of the perturbative expansion. The perturbative expansion for $U_I$ is
\beq
U_I = 1 + \frac{i}{2} \int_0^t ds_1 ~\lambda_0(s_1) \varphi_0^2(s_1) + \left(\frac{i}{2} \right)^2 \int_0^{t} ds_1 \int_0^{s_1} ds_2 ~\lambda_0(s_1) \varphi_0^2(s_1) \lambda_0(s_2) \varphi_0^2(s_2) + ...\nonumber\\
\eeq
where we have been careful with the time ordering. The perturbative expansion for $U_I^\dagger$ is
\beq
U^\dagger_I = 1 - \frac{i}{2} \int_0^t ds_1~ \lambda_0(s_1) \varphi_0^2(s_1) + \left(\frac{-i}{2} \right)^2 \int_0^{t} ds_1 \int_0^{s_1} ds_2 ~\lambda_0(s_2) \varphi_0^2(s_2) \lambda_0(s_1) \varphi_0^2(s_1) + ... \nonumber\\
\eeq
where we have been careful to reverse the terms in the second order expression. We also used $\lambda_0^\dagger = \lambda_0$ and $\varphi_0^\dagger = \varphi_0$ and have dropped the spatial arguments in the expressions above.

The leading contribution to $\mathcal{C}$ is obtained by setting $U_I = 1$. Then since
\beq
[\varphi_0(\x,t),\varphi_0(0)] = i \CG_R(\x,t)
\eeq
is proportional to the identity operator, we obtain
\beq
\mathcal{C}_0(\x,t) = - (i \CG_R)^2 = \CG_R^2(\x,t).
\eeq

The first order contribution is obtained by expanding $U_I$ to first order,
\beq
U_I^\dagger \varphi_0(\x,t) U_I = \varphi_0(\x,t) + \frac{i}{2} \int_0^t ds \int_{\y} \left[\varphi_0(\x,t), \lambda_0(\y,s) \varphi_0^2(\y,s) \right] + ...
\eeq
Since we need two powers of $\lambda_0$ to get a non-zero thermal expectation value (the expectation value giving the mass has been subtracted out), we need to either expand $U_I$ on a single time fold to higher order or expand $U_I$ to first order on both time folds. The first option leads to the standard self-energy correction along a single time fold and will not be reviewed here. The second option gives the first non-trivial ladder contribution. We have
\begin{eqnarray}
  \mathcal{C}_1 \sim - \tr&\bigg\{&\rho^{1/2}\left[ \frac{i}{2 } \int_0^t ds \int_{\y} \left[\varphi_0(\x,t),\lambda_0(\y,s) \varphi_0^2(\y,s) \right], \varphi_0(0) \right] \nonumber \\
   & \times &  \rho^{1/2} \left[ \frac{i}{2} \int_0^t ds' \int_{\y'} \left[\varphi_0(\x,t),\lambda_0(\y',s') \varphi_0^2(\y',s') \right], \varphi_0(0) \right] \bigg\}.
\end{eqnarray}
Using the free particle results $[\lambda_0,\varphi_0]=0$, $[\varphi_0(\x,t),\varphi_0(0)] = i \CG_R(\x,t)$, and $[\varphi_0(\x,t),\varphi_0^2(\y,s)] = [i \CG_R(\x-\y,t-s)] 2 \varphi_0(\y,s)$ (valid provided $t>s$) gives,
\beq
\mathcal{C}_1 = - i^2 \int_{s,s'} \int_{\y,\y'} [i \CG_R(\x-\y,t-s)][i \CG_R(\y,s)] [i \CG_R(\x-\y',t-s')]\nonumber\\
~~~~~~\times[i \CG_R(\y',s')] \CG_{\lambda,W}(\y-\y',s-s')
\eeq
or
\beq
\mathcal{C}_1 = \int_{s,s'} \int_{\y,\y'} \CG_R(\x-\y,t-s)\CG_R(\y,s)\CG_R(\x-\y',t-s')\CG_R(\y',s')\CG_{\lambda,W}(\y-\y',s-s') .\nonumber\\
\eeq
Up to factors of $1/N$ this is precisely the claimed one rung contribution of the type-I rung. Note the absence of any extra combinatorial prefactors.

The two rung contribution to the ladder is obtained by expanding $U_I$ to second order on both time folds. Suppressing the spatial labels for ease of notation, the second order expansion is
\begin{eqnarray}
  \left[ U_I^\dagger \varphi_0(t) U_I \right]_{2}& = & \frac{(-i)^2}{4}\int_0^{t} ds_1 \int_0^{s_1} ds_2 ~\lambda_0(s_2)\varphi_0^2(s_2) \lambda_0(s_1)\varphi_0^2(s_1) \varphi_0(t) \nonumber \\
    &+& \frac{i(-i)}{4} \int_0^{t} ds_1 \int_0^t ds_2 ~\lambda_0(s_1)\varphi_0^2(s_1) \varphi_0(t) \lambda_0(s_2)\varphi_0^2(s_2) \nonumber \\
    &+& \frac{i^2}{4} \int_0^{t} ds_1 \int_0^{s_1} ds_2 ~ \varphi_0(t) \lambda_0(s_1)\varphi_0^2(s_1) \lambda_0(s_2)\varphi_0^2(s_2).
\end{eqnarray}
Extracting an overall factor of $i^2/4$ and breaking the $s_1$ integral in the second term into two pieces ($s_2 < s_1$ and $s_2 > s_1$) and relabelling time integration variables, we recognize the formula
\beq
\left[ U_I^\dagger \varphi_0(t) U_I \right]_{2} = \left(\frac{i}{2} \right)^2 \int_0^{t} ds_1 \int_0^{s_1} ds_2 \left[ \left[ \varphi_0(t), \lambda_0(s_1) \varphi_0^2(s_1) \right], \lambda_0(s_2) \varphi_0^2(s_2) \right].
\eeq

There are now two types of terms that arise from the nested commutators. The first type generates the two-rung contributions from the type-I rung and the second type is responsible for the one-rung contribution from the type-II rung (see Figure \ref{laddersum}). The expansion of the nested commutator from above is,
\beq
[[,],] = [[\varphi_0(t),\varphi_0(s_1)] 2 \lambda_0(s_1) \varphi_0(s_1), \lambda_0(s_2) \varphi_0^2(s_2)]
\eeq
which gives
\begin{eqnarray} \label{2ndcomm}
  [[,],] &=& 4 [\varphi_0(t), \varphi_0(s_1)] [\varphi_0(s_1),\varphi_0(s_2)] \lambda_0(s_1) \lambda_0(s_2) \varphi_0(s_2) \nonumber \\
   &+& 2 [\varphi_0(t), \varphi_0(s_1)] [\lambda_0(s_1), \lambda_0(s_2)] \varphi_0(s_1) \varphi_0^2(s_2).
\end{eqnarray}
As indicated above, the first line of Eq.~\eqref{2ndcomm} gives rise to a two-rung contribution with type-I rungs as well as a self-energy correction for $\varphi$ (depending on whether the $\lambda$ fields are contracted within a time fold or between time folds). The second line of Eq.~\eqref{2ndcomm} gives rise to a one-rung contribution with a type-II rung and a self-energy correction for $\lambda$ (coming from the two terms in $[\varphi_0(s_1)\varphi_0^2(s_2), \varphi_0(0)]$).

Notice that for the two-rung contribution, the factors of $4$ from the vertex and nested commutators cancel. The type-II rung contribution arises from the case where $\varphi_0^2(s_2)$ commutes with $\varphi_0(0)$ to give $2 \varphi_0(s_2) [\varphi_0(s_2) , \varphi_0(0)]$ and the remaining $\varphi_0(s_1) \varphi_0(s_2)$ factor is contracted with a corresponding pair on the other time fold. For this type two rung, the factors of $4$ also cancel. It remains to establish the overall sign. The expression is
\begin{eqnarray}
  \mathcal{C}_2 &=& - i^4 \int_0^t ds_1 \int_0^{s_1} ds_2 \int_0^t ds'_1 \int_0^{s'_1} ds'_2 ~ [i \CG_R(t-s_1)] [i \CG_{\lambda,R}(s_1-s_2)] [i \CG_R(s_2)] \nonumber \\
   &\times& \CG_W(s_1 - s'_1) \CG_W(s_2 - s'_2) [i \CG_R(t-s'_1)] [i \CG_{\lambda,R}(s'_1-s'_2)] [i \CG_R(s'_2)]
\end{eqnarray}
or
\begin{eqnarray}
  \mathcal{C}_2 &=& \int_0^t ds_1 \int_0^{s_1} ds_2 \int_0^t ds'_1 \int_0^{s'_1} ds'_2 ~  \CG_R(t-s_1) \CG_{\lambda,R}(s_1-s_2) \CG_R(s_2) \nonumber \\
   &\times& \CG_W(s_1 - s'_1) \CG_W(s_2 - s'_2) \CG_R(t-s'_1) \CG_{\lambda,R}(s'_1-s'_2) \CG_R(s'_2).
\end{eqnarray}
Again, this is the claimed one rung contribution of type two including the overall positive sign and combinatorial factor of unity.

Following the manipulations for the first and second order terms, the general term starts from the expansion
\beq
\left[ U_I^\dagger \varphi_0(t) U_I \right]_{n} = \left(\frac{i}{2}\right)^n \int_0^{t} ds_1 ... \int_0^{s_{n-1}} ds_n \underbrace{\left[ ... \left[ \varphi_0(t), \lambda_0(s_1) \varphi_0^2(s_1) \right] ... ,  \lambda_0(s_n) \varphi_0^2(s_n) \right]}_{\text{$n$ nested commutators}}.
\eeq
The various multi-rung ladder diagrams all arise from these nested commutators. The following claims can be straightforwardly verified from the general expansion. First, the combinatorial factors always cancel to give unity in ladder diagrams with either rung type. Second, the time ordering guarantees that every commutator can be replaced by a retarded Green's function; furthermore, including the explicit factor of $\theta(t-s)$ in $G_R(t-s)$ allows us to extend all time integrals to run over the entire line $(-\infty,\infty)$. Third, the overall sign is positive when written as an integral over products of $\CG_R$ and $\CG_W$. This follows because in an $\ell$ rung ladder, there are $2\ell +2$ commutators which each give a factor of $i$, there are $2\ell$ factors of $i$ from the interaction vertices, and there is an overall minus sign, which gives a total sign of $-i^{2\ell+2} i^{2\ell} = +1$.

\section{Ladder versus crossed diagrams}
\label{cross}
\begin{figure}
  \centering
  \includegraphics[width=.6\textwidth]{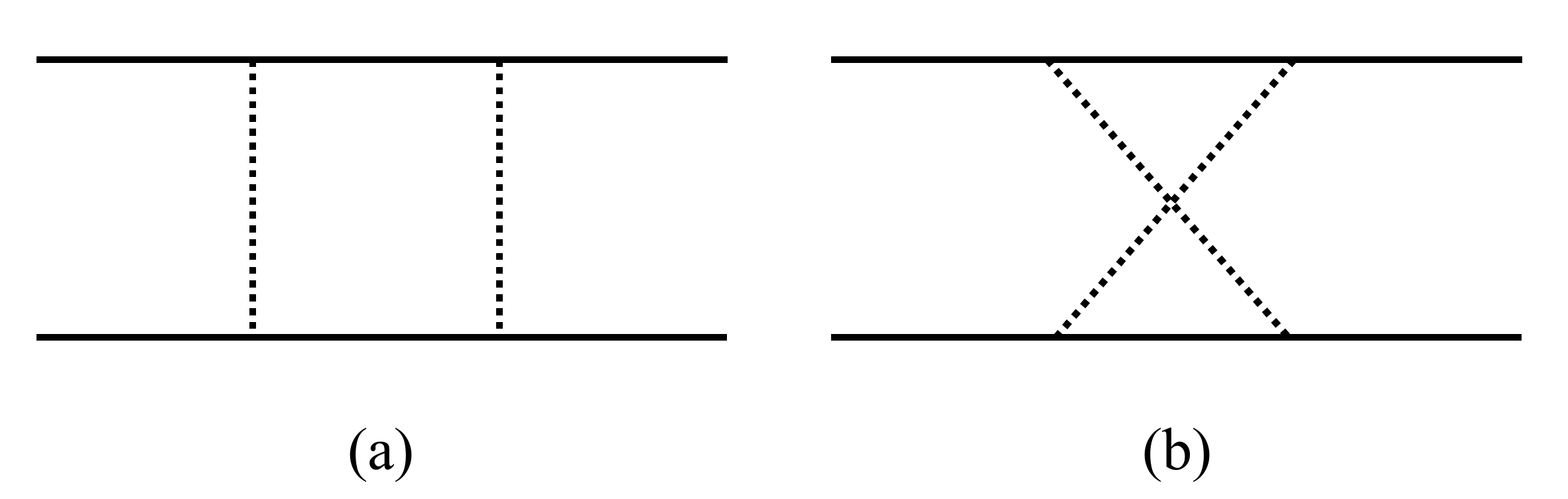}
  \caption{(a) $\chi_{\tn{ladder}}$, (b) $\chi_{\tn{crossed}}$. Here the rungs could be either type-I or type-II.}
  \label{ladcross}
\end{figure}

In this appendix, we discuss an important issue, namely whether it is justified to retain only the ladder series (see Figure~\ref{ladcross}a) as discussed earlier or if one needs to include the crossed diagrams as shown in Figure~\ref{ladcross}b. In order to study this aspect in more detail, consider the two-rung ladder and two-rung crossed diagrams respectively, and focus only on the internal ``loop".

The internal loop in the two-rung ladder is given by,
\beq
\chi_{\tn{ladder}} &=& \int\frac{d\ve}{2\pi}~\int_{\l}~\int\frac{d\omega}{2\pi}~\int_{\p}~\int\frac{d\omega'}{2\pi}~\int_{\p'}~ ~\CG_R(\nu-\ve,-\l)~\CG_R(\ve,\l)~\CG_{W,\lambda}(\ve-\omega,\l-\p)\nonumber\\
&&\CG_R(\nu-\omega,-\p)~\CG_R(\omega,\p)~\CG_{W,\lambda}(\omega-\omega',\p-\p')~\CG_R(\nu-\omega',-\p')~\CG_R(\omega',\p').
\eeq

Similarly, the loop in the two-rung crossed diagram is given by,
\beq
\chi_{\tn{crossed}} &=& \int\frac{d\ve}{2\pi}~\int_{\l}~\int\frac{d\omega}{2\pi}~\int_{\p}~\int\frac{d\omega'}{2\pi}~\int_{\p'}~\CG_R(\nu-\ve,-\l)~\CG_R(\ve,\l)~\CG_{W,\lambda}(\ve-\omega,\l-\p)\nonumber\\
&&\CG_R(\nu+\omega-\omega'-\ve,\p-\p'-\l)~\CG_R(\omega,\p)~\CG_{W,\lambda}(\omega-\omega',\p-\p')\nonumber\\
&&~~~~~~~~~~~~~~~~~~~~~\CG_R(\nu-\omega',-\p')~\CG_R(\omega',\p').
\eeq

Let us now compare and contrast the pole structures of the above diagrams. For the ladder diagrams, the retarded propagators always occur in pairs of the form, $\CG_R(\nu-\omega,-\p)~\CG_R(\omega,\p)$, with an integral over $\p$, which upon doing the $\omega$ integral leads to a pole structure dominated by the $\nu^{-1}$ term, as far as the long time form of $\C(t)$ is concerned. Moreover, the contribution to $\chi_{\tn{ladder}}$ above from all three pieces then goes as $\nu^{-3}$, since the different pairs of the retarded propagators depend only on one frequency that is being integrated over (e.g. $\ve, \omega, \omega'$) {\footnote{The frequencies are, of course, not entirely decoupled. They are coupled via the rungs of the ladder, $\CG_{W,\lambda}$.}}.

On the other hand, notice that for the crossed diagram, the combination of the retarded propagators coming from the ``cross" (or the flipped; see fig. \ref{ladcross}b) region is,
\beq
\CG_R(\nu+\omega-\omega'-\ve,\p-\p'-\l)~\CG_R(\omega,\p).
\eeq
Note that the momenta in the two legs are not directed opposite to each other. Carrying out the $\omega$ integral leads to pole structures of the form: $(\nu-\omega'-\ve \pm \epsilon_{\p} \pm \epsilon_{\p-\p'-\l})^{-1}$. This can be combined with the contributions from the other pieces in $\chi_{\tn{crossed}}$. It is not hard to see that upon doing the $\omega'$ integral leads to a pole structure: $\left[\nu (\nu-\ve \pm \epsilon_{\p'} \pm \epsilon_{\p} \pm \epsilon_{\p-\p'-\l}) \right]^{-1}$. Finally, doing the $\ve$ integral leads to: $\left[\nu^2 (\nu \pm \epsilon_{\l} \pm \epsilon_{\p'} \pm \epsilon_{\p} \pm \epsilon_{\p-\p'-\l}) \right]^{-1}$.

There is an equivalent real time picture of the suppression of the crossed ladder relative to the uncrossed ladder. Taking for simplicity two type-I rungs, the uncrossed ladder is schematically
\begin{align}
  \chi_{\text{ladder}} \sim \int_{s_1,s_2,s'_1,s'_2} &\CG_R(t-s_1) \CG_R(s_1 - s_2) \CG_R(s_2) \CG_{W,\lambda}(s_1 - s'_1)\CG_{W,\lambda}(s_2 - s'_2) \nonumber \\
   &\times  \CG_R(t-s'_1) \CG_R(s'_1 - s'_2) \CG_R(s'_2) .
\end{align}
Roughly speaking, the effect of the Wightman propagators is to set $s_i = s'_i$. Said differently, the Wightman functions constrain the relative variables $s_i - s'_i \approx 0 $ to a precision of about $\beta$ but the center of mass variables $\frac{s_i + s'_i}{2}$ remain unconstrained. Then the ladder contribution is schematically
\beq
\chi_{\text{ladder}} \sim \beta^2 \int_{s_1,s_2} [\CG_R(t-s_1)]^2 [\CG_R(s_1 - s_2)]^2 [\CG_R(s_2)]^2.
\eeq

On the other hand, the crossed ladder is schematically
\begin{align}
  \chi_{\text{crossed}} \sim \int_{s_1,s_2,s'_1,s'_2} &\CG_R(t-s_1) \CG_R(s_1 - s_2) \CG_R(s_2) \CG_{W,\lambda}(s_1 - s'_2)\CG_{W,\lambda}(s_2 - s'_1) \nonumber \\
   &\times  \CG_R(t-s'_1) \CG_R(s'_1 - s'_2) \CG_R(s'_2) .
\end{align}
The Wightman functions now constrain $s_1 - s'_2 \approx 0$ and $s_2 - s'_1 \approx 0$. However, since the integrand is zero unless $s_1 > s_2$ and $s'_1 > s'_2$, the effective range of integration is severely restricted by these constraints. We may write the causality constraints as
\beq
\theta(s_1 -s_2 ) \theta(s'_1 - s'_2) \approx \theta(s_1 - s_2) \theta(s_2 - s_1) ,
\eeq
so we also have $s_1 - s_2 \approx 0$ to a precision of about $\beta$. This further restriction of the time arguments in the integral implies that $\chi_{\text{crossed}}$ is parametrically smaller than $\chi_{\text{ladder}}$.

As a simple model of this suppression, consider a free particle with position $X$ and decay rate $\gamma$. The equation of motion is $ \partial_t^2 X = - \gamma \partial_t X$. The retarded correlator is
\beq
\CG_{R,x}(t) \propto \theta(t) (1 - e^{- \gamma t}).
\eeq
For $\gamma t \gg 1$ and $\gamma \beta \ll 1$ the ratio of the ladder to the crossed ladder is then of order
\beq
\frac{\chi_{\text{ladder}}}{\chi_{\text{crossed}}}\bigg|_{\text{particle model}} \sim \frac{t^2}{\beta t (\beta \gamma)^2} \sim  \frac{t}{\beta^3 \gamma^2} \gg 1.
\eeq

\section{Details of the numerical computations}
\label{sec:numerics}

Here we discuss the details of our numerical computation of $\lambda_L$ and $v_B$. The first step is to compute the analytic continuation of the bubble. The imaginary part is computed by numerical integration of Eq.~\eqref{imbformula} using a standard MATLAB routine. The real part is then obtained by a stable version of Kramers-Kronig,
\beq
\text{Re}[\Pi_R(\omega,\q)] = \frac{2}{\pi} \int_0^\infty d\nu \frac{\omega \text{Im}[\Pi_R(\omega,\q)] - \nu \text{Im}[\Pi_R(\nu,\q)] }{\omega^2 - \nu^2}.
\eeq
Here we have subtracted a multiple of the zero integral $\mathbb{P} \int d\nu  \frac{1}{\omega^2 -\nu^2} = 0$ so that the resulting integrand is non-singular at $\omega = \nu$ and the principal part can be removed.

This integral is computed in three parts because the imaginary part is non-analytic at $\omega = |\q|$ and $\omega = \sqrt{|\q|^2 + 4\mu^2}$. The numerical integral defining the real part is then carried out by breaking it into three pieces, $\nu \in (0,|\q|)$, $\nu \in (|\q|,\sqrt{\q^2+4\mu^2})$, and $\nu \in (\sqrt{\q^2+4\mu^2},\infty)$. The upper limit is taken to be a large number of order $10^4$ so that the result is converged to several digits. We have also checked the result by treating the discontinuity in $\text{Im}[\Pi_R]$ at $\omega = \sqrt{\q^2 + 4\mu^2}$ exactly and performing Kramers-Kronig on the continuous subtracted imaginary part. The methods agree to several digits.

The values of the real and imaginary parts are precomputed on a relatively fine grid in frequency and momentum for later use (this precomputing is necessary to avoid prohibitive slowdowns in later steps). For the results quoted in the main text, the precomputed grid is $400 \times 400$ with $\omega/T, |\q|/T \in [0,12]$.

Given the real and imaginary parts of the bubble, we are then able to compute the $\lambda$ spectral function $A_\lambda$, the $\lambda$ retarded correlation function $\CG_{\lambda,R}$, and the $\lambda$ Wightman function $\CG_{\lambda,W}$. If values of $\Pi_R(|\vec{q}|,\omega)$ not in the precomputed grid are needed, then we use two dimensional spline interpolation to determine approximate values. If $|\vec{q}|$ or $\omega$ lie outside the grid, then we set the value of $\Pi_R$ to zero. These are additional approximations.

From here the calculation follows two routes depending on whether we are interested in the $\vec{k}=0$ chaos exponent or the $\vec{k}\neq 0$ chaos exponent.

\subsection{$\vec{k}=0$}

The basic assumption underlying this calculation is that the growing mode is spherically symmetric and depends only on $|\vec{p}|$. Hence we use radial coordinates so that the angular variable drops out of the final eigenvalue equation.

Using the computed values of the propagators, we first obtain the dephasing rate $\Gamma_{\q}$ by numerically integrating the rung function $\Ru_1$ over angle and then over radial momentum. The angular grid is taken be uniform with $n_\theta = 60$ points, while the radial grid is also uniform and has up to $n_{\text{1d}} = 200$ points.

We then average $\Ru_1(\vec{\ell},\vec{p})$ and $\Ru_2(\vec{\ell},\vec{p})$ over the angle between $\vec{\ell}$ and $\vec{p}$ for $|\vec{\ell}|$ and $|\vec{p}|$ on the radial grid. Note that there is an extra complication here, since the calculation of $\Ru_2(\vec{\ell},\vec{p})$ already involves an angular integral coming from the integral over the loop momentum. This angular integral is also done approximately on a grid with $n_\theta = 60$ points.

Finally, using the dephasing rate computed before and the angular averaged versions of $\Ru_1$ and $\Ru_2$, we numerically solve a discretized version of the integral equation Eq.~\eqref{rung3}. We generically find one or a few real positive eigenvalues and then many more complex eigenvalues with negative real parts. By studying the system size dependence of the results, we attempt to approximately extrapolate to a converged limit.

\subsection{$\vec{k}\neq 0$}

In this version of the calculation we cannot assume that the growing mode is spherically symmetric. Hence we work on a general two-dimensional momentum grid. Another possibility is to break the equation down into partial waves and solve a set of coupled radial integral equations, but we did not pursue it. We define a square grid of vectors $(q_x,q_y)$ with $|q_x|,|q_y| < k_{\max}$ and $n_{\text{2d}}^2$ points. We consider up to $n_{\text{2d}} = 61$ in our calculation, corresponding to matrices of size $n_{\text{2d}}^2 \times n^2_{\text{2d}} = 3721 \times 3721$. Since the matrix is not sparse, memory sets a limit to how fine a grid we can consider. However, the time needed to set all the values of the matrices is by far the limiting step.

\begin{figure}
  \centering
  \includegraphics[width=.8\textwidth]{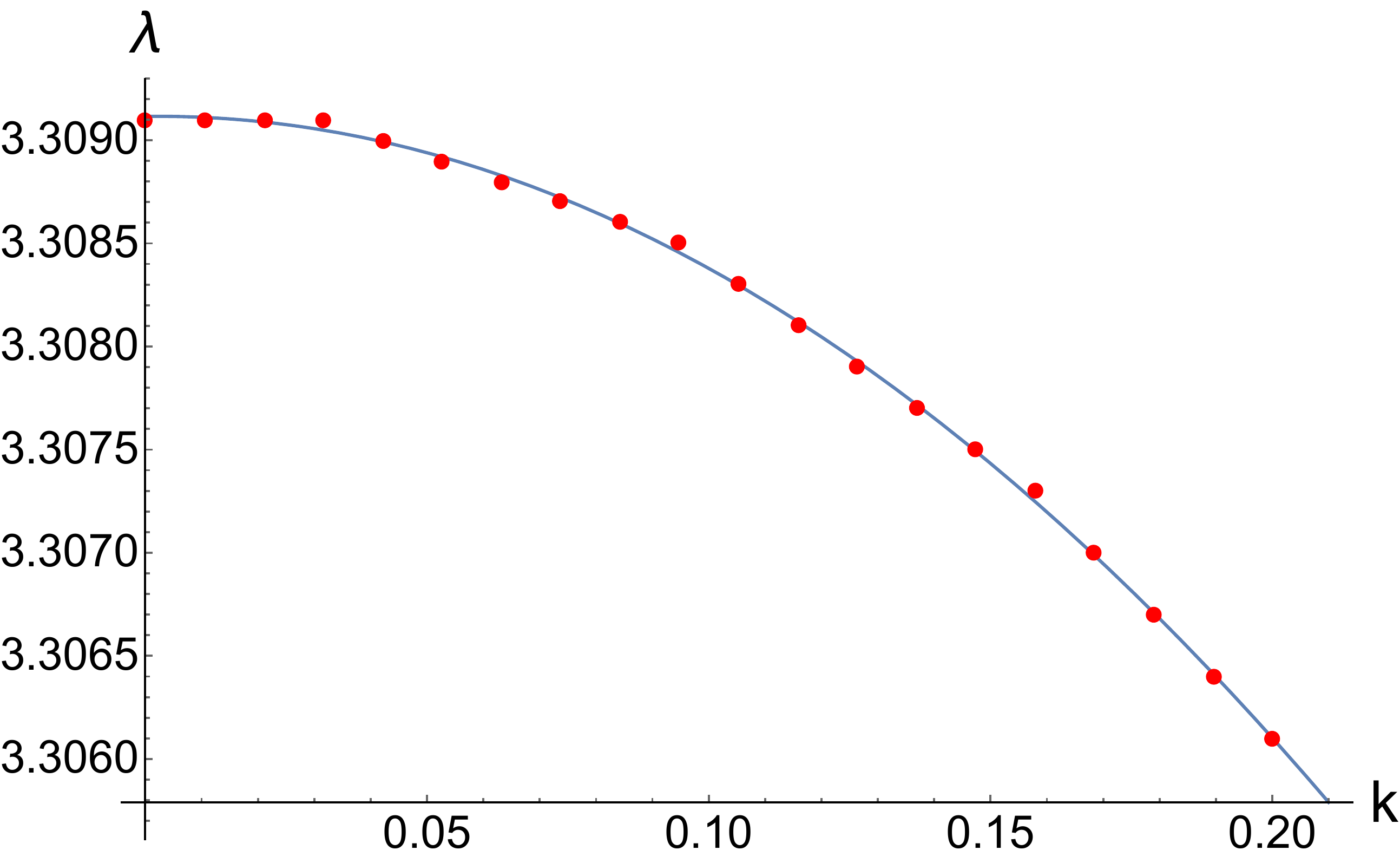}
  \caption{Dependence of chaos exponent on scaled external momentum for $n_{\text{2d}} = 51$. The quadratic fit to the data is shown in blue.}
  \label{fig:Lvsk}
\end{figure}

The dephasing rate is again computed by numerically integrating the formula for $\Gamma_{\q}$ over the two-dimensional grid. The value of $\Ru_1$ and $\Ru_2$ are also computed on the grid and the additional angular integral entering the definition of $\Ru_2$ is done approximately using $n_\theta = 60$. We then solve the eigenvalue equation as a function of external momentum $\vec{k}$. To be precise, as discussed in the main test, by working with the scaled variable $N \vec{k}/T$, the entire eigenvalue equation is rendered proportional to $T/N$. We find that the leading eigenvalue is positive and real and obeys
\beq
\lambda(\vec{k}) = \lambda_0 - \lambda_2 |\vec{k}|^2 + ...
\eeq
for $N \vec{k}/T \ll 1$. This is shown in Figure \ref{fig:Lvsk}. There are also subleading eigenvalues including complex eigenvalues with negative real parts. Again, by studying the system size dependence of the results, we attempt to approximately extrapolate to a converged limit.

\subsection{Run times}

With the $400 \times 400$ $\omega,|\q|$ grid, the full precomputation step including the calculation of the type-II rung takes a few days in MATLAB running on a laptop with two $2.4$ GHz cores and $8$ GB of memory. The calculation of the rungs in the spherically symmetric case is comparatively quick, with the matrix element setup and subsequent diagonalization taking about an hour for the largest grid sizes considered ($n_{\text{1d}} = 200$). By contrast, the matrices involved in the calculation of $v_B$ are much larger and take much longer to initialize. The matrix element setup and subsequent diagonalization took about a day and a half for the largest grid sizes considered ($n_{\text{2d}} = 61$). All told, the setup and data collection took about a week for the data presented in the main text.

As indicated above, the most expensive step by far is the initialization of the matrices. The largest matrices considered were less than $4000 \times 4000$, so memory was not the primary limitation in our calculation. Given more time and further optimizations of the code and possibly parallelization, the calculation could be further improved as needed.

\subsection{Extrapolation in to infinite $\omega,|\q|$ grid size}

In the main text we reported results for the extrapolation of $1/\tau_\phi$, $\lambda_L$, and $v_B$ to infinite grid size in the solution of the integral equation. However, there are two additional grid sizes in the problem, the initial fine grid in $\omega,|\q|$ and the angular grid used to approximately compute the type-II rung. The results did not appear to depend too sensitively on the angular grid size, but because of limited time we did not pursue a systematic study of this dependence.

We did study the results as a function of the precomputed grid size. We repeated the calculations in the main text for a precomputed grid size of $200 \times 200$. We then took the extrapolated values for $400$ and $200$ and performed a further linear fit to one over the precomputed grid size. Of course this is only two data points and we have not carefully studied the dependence on the angular grid, so our conclusions in this appendix are tentative.

The estimated infinite grid size values for the $\k =0$ calculation and the $\k \neq 0$ calculation are
\begin{align}
\frac{1}{\tau_\phi}\bigg|_{\k=0 \,\text{calc}} &= 1.12 \nonumber \\
\lambda_{L}\big|_{\k=0 \,\text{calc}} &= 3.15 \nonumber  \\
\frac{1}{\tau_\phi}\bigg|_{\k\neq0 \,\text{calc}} &= 1.21 \nonumber  \\
\lambda_{L}\big|_{\k\neq0 \,\text{calc}} &= 3.12 \nonumber  \\
v_B \big|_{\k \neq 0 \,\text{calc}} &= 1.01.
\end{align}
The agreement between the two methods is reasonable. Comparison with the high precision value of $1/\tau_\phi$ suggests errors at the level of at least $5$\%. Obviously the butterfly velocity must obey $v_B \leq 1$, although the extrapolation gives a value slightly larger than one. Within the accuracy of the calculation we simply say that $v_B \approx 1$.
\bibliographystyle{JHEP}
\bibliography{scrambling}
\end{document}